\RequirePackage{lineno}
\documentclass[aps,prc,preprint,superscriptaddress,floatfix]{revtex4-1}

\usepackage{amsmath}
\usepackage[english]{babel}
\usepackage[bookmarks=false,colorlinks=false]{hyperref}
\usepackage{graphicx}
\usepackage{url}
 \usepackage{footnote}

\begin{document}


\title{ Phase-space dependence of particle-ratio fluctuations in Pb+Pb collisions from 20\textit{A} to 158\textit{A} GeV beam energy}

\author{T.~Anticic$^{21}$, B.~Baatar$^{8}$, D.~Barna$^{4}$, J.~Bartke$^{6}$, 
H.~Beck$^{9}$, L.~Betev$^{10}$, H.~Bia{\l}\-kowska$^{18}$, C.~Blume$^{9}$, 
B.~Boimska$^{18}$, J.~Book$^{9}$, M.~Botje$^{1}$,
P.~Bun\v{c}i\'{c}$^{10}$,
P.~Christakoglou$^{1}$,
P.~Chung$^{17}$, O.~Chvala$^{14}$, J.~Cramer$^{15}$, V.~Eckardt$^{13}$,
Z.~Fodor$^{4}$, P.~Foka$^{7}$, V.~Friese$^{7}$,
M.~Ga\'zdzicki$^{9,11}$, K.~Grebieszkow$^{20}$, C.~H\"{o}hne$^{7}$,
K.~Kadija$^{21}$, A.~Karev$^{10}$, V.~Kolesnikov$^{8}$, M.~Kowalski$^{6}$, 
D.~Kresan$^{7}$,
A.~Laszlo$^{4}$, R.~Lacey$^{17}$, M.~van~Leeuwen$^{1}$,
M.~Ma\'{c}kowiak-Paw{\l}owska$^{9,20}$, M.~Makariev$^{16}$, A.~Malakhov$^{8}$,
G.~Melkumov$^{8}$, M.~Mitrovski$^{9}$, S.~Mr\'owczy\'nski$^{11}$, 
G.~P\'{a}lla$^{4}$, A.~Panagiotou$^{2}$, W.~Peryt$^{20}$, 
J.~Pluta$^{20}$, D.~Prindle$^{15}$,
F.~P\"{u}hlhofer$^{12}$, R.~Renfordt$^{9}$, C.~Roland$^{5}$, G.~Roland$^{5}$,
M. Rybczy\'nski$^{11}$, A.~Rybicki$^{6}$, A.~Sandoval$^{7}$, 
A. Rustamov$^{9}$,
N.~Schmitz$^{13}$, T.~Schuster$^{9}$, P.~Seyboth$^{13}$, F.~Sikl\'{e}r$^{4}$, 
E.~Skrzypczak$^{19}$, M.~S{\l}\-odkowski$^{20}$, G.~Stefanek$^{11}$, R.~Stock$^{9}$, 
H.~Str\"{o}bele$^{9}$, T.~Susa$^{21}$, M.~Szuba$^{20}$, 
D.~Varga$^{3}$, M.~Vassiliou$^{2}$,
G.~Veres$^{4}$, G.~Vesztergombi$^{4}$, D.~Vrani\'{c}$^{7}$,
Z.~W{\l}odarczyk$^{11}$, A.~Wojtaszek-Szwarc$^{1}$}

\vspace{0.5cm}

\affiliation{
\mbox{$^{}$ NIKHEF, Amsterdam, Netherlands.}\\
\mbox{$^{2}$ Department of Physics, University of Athens, Athens, Greece.}\\
\mbox{$^{3}$ E\"otv\"os Lor\'ant University, Budapest, Hungary.}\\
\mbox{$^{4}$ Wigner Research Center for Physics, Hungarian Academy of Sciences, Budapest, Hungary.}\\
\mbox{$^{5}$ MIT, Cambridge, Massachusetts, USA.}\\
\mbox{$^{6}$ H.~Niewodnicza\'nski Institute of Nuclear Physics, Polish Academy of Sciences, Cracow, Poland.}\\
\mbox{$^{7}$ GSI Helmholtzzentrum f\"{u}r Schwerionenforschung GmbH, Darmstadt, Germany.}\\
\mbox{$^{8}$ Joint Institute for Nuclear Research, Dubna, Russia.}\\
\mbox{$^{9}$ Fachbereich Physik der Universit\"{a}t, Frankfurt, Germany.}\\
\mbox{$^{10}$ CERN, Geneva, Switzerland.}\\
\mbox{$^{11}$ Institute of Physics, Jan Kochanowski University, Kielce, Poland.}\\
\mbox{$^{12}$ Fachbereich Physik der Universit\"{a}t, Marburg, Germany.}\\
\mbox{$^{13}$ Max-Planck-Institut f\"{u}r Physik, Munich, Germany.}\\
\mbox{$^{14}$ Institute of Particle and Nuclear Physics, Charles University, Prague, Czech Republic.}\\
\mbox{$^{15}$ Nuclear Physics Laboratory, University of Washington, Seattle, Washington, USA.}\\
\mbox{$^{16}$ Institute for Nuclear Research and Nuclear Energy, BAS, Sofia, Bulgaria.}\\
\mbox{$^{17}$ Department of Chemistry, Stony Brook University (SUNYSB), Stony Brook, New York, USA.}\\
\mbox{$^{18}$ National Center for Nuclear Research, Warsaw, Poland.}\\
\mbox{$^{19}$ Institute for Experimental Physics, University of Warsaw, Warsaw, Poland.}\\
\mbox{$^{20}$ Faculty of Physics, Warsaw University of Technology, Warsaw, Poland.}\\
\mbox{$^{21}$ Rudjer Boskovic Institute, Zagreb, Croatia.}\\
}

\thanks{Corresponding author: a.rustamov@cern.ch}



\begin{abstract}

A novel approach, the identity method, was used for particle identification
and the study of fluctuations of particle yield ratios in Pb+Pb collisions
at the CERN Super Proton Synchrotron (SPS). This procedure allows to unfold the moments of the unknown
multiplicity distributions of protons (p), kaons (K), pions ($\pi$) and electrons (e).
Using these moments the excitation function of the fluctuation measure $\nu_{\text{\text{dyn}}}$[A,B]
was measured, with A and B denoting different particle types.
The obtained energy dependence of $\nu_{\text{dyn}}$ agrees with previously published NA49
results on the related measure $\sigma_{\text{dyn}}$. Moreover, $\nu_{\text{dyn}}$ was found
to depend on the phase space coverage for [K,p] and [K,$\pi$] pairs.
This feature most likely explains the reported differences between measurements
of NA49 and those of STAR in central Au+Au collisions.


\end{abstract}

\maketitle

\section{Introduction}
\label{intro}
By colliding heavy ions at high energies one hopes to heat and/or compress the matter to energy densities 
at which the production of the Quark-Gluon Plasma (QGP) begins~\cite{QGP1,QGP2}. 
Lattice QCD calculations can study this
non-perturbative regime of QCD~\cite{LatticeQCD} and allow a quantitative investigation of the QGP properties.
A first order phase boundary is expected to separate high temperature hadron matter from the QGP 
for large net baryon density and is believed to end in a critical point~\cite{Stephanov}. 
A wealth of ideas have been proposed to explore the 
properties and the phase structure of strongly interacting matter.  Event-by-event fluctuations of 
various observables may be sensitive to the transitions between hadronic and partonic phases~\cite{StepRajShur,Koch}. 
Moreover, the location of the critical point may be signalled by a characteristic pattern in the energy 
and system size dependence of the measured fluctuation signals. 

Pb+Pb reactions were investigated at the CERN SPS since 1994 by a variety of experiments at the top SPS energy. Many of
the predicted signals of the QGP were observed~\cite{HeinzJacob}, but their uniqueness was in doubt. Motivated
by predictions of the Statistical Model for the Early Stage of nucleus-nucleus collisions~\cite{SMES} of characteristic changes of hadron production properties at
the onset of QGP creation (onset of the deconfinement) the NA49 experiment performed a scan of the entire 
SPS energy range, from 158\textit{A} down to 20\textit{A} GeV. The predicted features were found at an
energy of about 30\textit{A} GeV in central Pb+Pb collisions~\cite{NA49_horn}, thereby indicating the onset
of deconfinement in collisions of heavy nuclei in the SPS beam energy range. 
These observations have recently been confirmed by the RHIC beam energy scan and the 
expected trend towards higher energy is consistent with LHC data ~\cite{NA49_Rustamov}. 

Motivated by these findings the NA49 Collaboration has started to explore the phase diagram of
strongly interacting matter, with the aim of searching for indications of the first order phase transition 
and the critical point by studying several measures of fluctuations. In particular, the energy dependence 
of dynamical event-by-event fluctuations of the particle composition was investigated 
using the measure $\sigma_{\text{dyn}}$(A/B) with A and B denoting the multiplicities of different particle species. 
An increasing trend of $\sigma_{\text{dyn}}$ for both K/p and K/$\pi$ 
ratios towards lower collision energies was observed~\cite{NA49_fluct1,NA49_fluct2,NA49_fluct3}. In contrast, 
recent results of the STAR experiment from the Beam Energy Scan (BES) at the 
Relativistic Heavy Ion Collider (RHIC) show practically no energy dependence 
of the related event-by-event fluctuation measure $\nu_{\text{dyn}}$~\cite{voloshin_nu} 
for [K, p] and [K, $\pi$] pairs~\cite{STAR_fluct}. 
The comparison between NA49 and corresponding STAR results was performed using the relation

\begin{equation}
\label{relation}
\nu_{\text{dyn}} = \text{sgn}(\sigma_{\text{dyn}})\sigma_{\text{dyn}}^{2}.
\end{equation}

However, the accuracy of this relation decreases inversely with multiplicity, i.e. at lower energies this relation 
is only approximate. In order not to rely on this approximation the fluctuation measure $\nu_{\text{dyn}}$ 
was directly reconstructed in this paper using a novel identification scheme, the \textit{Identity Method}~\cite{identity1,identity2}.
The procedure avoids event-by-event particle ratio fits and the use of mixed events necessary to subtract the artificial
correlations introduced by the fits. Moreover, the much improved statistical power allows to study the effects
of the different phase space coverage of the NA49 (forward rapidities) and STAR (central rapidity, 
without low-$p_{\bot}$ range) experiments. 

The paper is organized as follows. Details about the detector setup and the data are given in section~\ref{secData}. 
Section~\ref{sec-cuts} discusses the event and track selection criteria. 
The novel features of this analysis, i.e. the particle identification procedure and the extraction of the moments
of the multiplicity distributions, are discussed in sections~\ref{sec-pid} and~\ref{sec-identity}, respectively.
Section~\ref{sec-errors} presents the estimates of statistical and systematic uncertainties.  
Results on $\nu_{\text{dyn}}$ and their phase-space dependence are discussed in sections ~\ref{sec-nu} and~\ref{sec-accepatance}. 
Finally, section~\ref{sec-summary} summarizes the paper.

\section{Experimental setup and the data}
\label{secData}
This paper presents results for central Pb+Pb collisions at projectile energies of 
20\textit{A}, 30\textit{A}, 40\textit{A}, 80\textit{A} and 158\textit{A} GeV, recorded 
by the NA49 experiment (for a detailed description of the NA49 apparatus cf. Ref.~\cite{NA49_NIM}). 
The principal tracking detectors are four large volume Time Projection Chambers  (TPC) with two of them, 
Vertex TPCs (VTPC1 and VTPC2), placed inside superconducting dipole magnets with a combined maximum bending 
power of 9 Tm for a length of 7 m.  Care was taken to keep the detector acceptance approximately
constant with respect to midrapidity by setting the magnetic field strength proportional to the
beam energy. Particle identification in this analysis is achieved 
by simultaneous measurement of particle momenta and their specific energy loss \textit{dE/dx} in the gas volume of 
the main TPCs (MTPC-L and MTPC-R). These are located downstream of the magnets on either side of the beam, 
have large dimensions (4 m $\times$ 4 m $\times$ 1.2 m) and feature 90 readout pad rows, providing an 
energy loss measurement with a resolution of about 4$\%$. In the experiment Pb beams with an intensity of 
10$^4$ ions/s were incident on a thin lead foil located 80 cm upstream of the VTPC-1. 
For 20\textit{A} - 80\textit{A}~GeV and 158\textit{A}~GeV the target thicknesses amounted to 0.224 g/cm$^{2}$ 
and 0.336 g/cm$^{2}$, correspondingly. The centrality of a collision was determined based on the energy of 
projectile spectators measured in the veto  calorimeter (VCAL) which is located 26 m behind the target 
and covers the projectile-spectator phase space region. A collimator in front of the calorimeter was 
adjusted for each energy in such a way that all projectile spectator protons, neutrons and beam fragments 
could reach the veto calorimeter while keeping the number of produced particles hitting the calorimeter 
as small as possible. 
   
 \begin{center}
 \begin{table*}
\begin{tabular}{|c|c|c|c|c|c|}
\hline 
\hline  Beam energy & $\sqrt{s_{NN}}$ & $N^{\text{events}}$ & $\langle N^{\text{all}} \rangle$ & $\langle N^{\text{pos.}} \rangle$ \\ [0.5ex] 
           [GeV] & [GeV] &  & &\\  [0.5ex] 
\hline  
20\textit{A} & 6.3 & 169k &63 & 46 \\  [0.5ex] 
\hline  
30\textit{A} & 7.6 & 179k & 113 & 75  \\  [0.5ex] 
\hline  
40\textit{A} & 8.7 & 195k & 159 & 99 \\  [0.5ex] 
\hline  
80\textit{A}& 12.3  & 136k &315 & 181 \\  [0.5ex] 
\hline  
158\textit{A}& 17.3 & 125k & 560 & 310 \\  [0.5ex]
 \hline
 \end{tabular}
\caption {The statistics corresponding to the 3.5$\%$ most central Pb+Pb collisions used in this analysis.}
\label{table1}
\end{table*}
\end{center}

\section{Event and track selection criteria}
\label{sec-cuts}
The only event selection criterion used in this analysis is  a centrality cut based on the energy (E$_{\text{Cal}}$) 
of forward going projectile spectators  measured in VCAL. 
The data were recorded with an online VCAL cut accepting the 7$\%$ and 10$\%$ most central Pb+Pb collisions
for 20\textit{A} - 80\textit{A}~GeV and 158\textit{A}~GeV, respectively. Using an offline cut on E$_{\text{Cal}}$,  
event samples of the 3.5$\%$ most central reactions were selected, which in the Glauber Monte Carlo Model corresponds 
to about 367 wounded nucleons and an impact parameter range of 0~$<$~b~$<$~2.8 fm~\cite{glauber}.
To ensure better particle separation only the tracks with large track length (better energy loss resolution) 
in the MTPCs were used for further analysis. For this purpose we distinguish between the number of potential 
and the number of reconstructed \textit{dE/dx} points. The former was estimated according to the position of the track 
in space together with the known TPC geometry,  while the latter represents the number of track points reconstructed 
by the cluster finder algorithm. In addition, to avoid the usage of track fragments (split tracks from different TPCs which were not matched together), it is required that more than
50 $\%$ of potential points have to be found by the reconstruction algorithm. The following track selection criteria, referred to as the "loose cuts", 
are used for the main analysis:
\begin{itemize}
\item The number of reconstructed points in the MTPCs should be more than 30.
\item The ratio of the number of reconstructed points in all TPCs (VTPCs + MTPCs) to the number of potential points in all TPCs should exceed 0.5.
\end{itemize}
These selections reduce the acceptance of the particles to the forward rapidity regions in the center-of-mass reference frame.
In order to study the systematic uncertainties of the final results due to the applied track cuts another set of cuts  ("tight cuts") 
was employed in addition to the "loose cuts":
\begin{itemize}
\item The number of potential points in at least one of the vertex TPCs (VTPC1 or VTPC2) and in the MTPCs  should be more than 10 and  30, respectively.
\item The ratio of the number of reconstructed points  to the number of potential points in the selected TPC(s) should exceed 0.5.
\item The distance between the closest point on the extrapolated track to the main vertex position should be less than 4 cm in $x$ 
(bending plane) and less than 2 cm in $y$ (vertical).
\end{itemize}

The statistics used in this analysis, with applied "loose cuts", is shown in Table~\ref{table1}. 

\section{Particle identification}
\label{sec-pid}
Particle identification (PID)  in this analysis is achieved by correlating the measured particle momentum with its 
specific energy loss \textit{dE/dx} in the gas volume of the MTPCs. The key problem of particle identification by \textit{dE/dx} measurement 
is the fluctuation of ionization losses. The energy loss distribution has a long tail for large values. 
Its shape was first calculated in 
Ref.~\cite{dedx_Landau} and is referred to as the Landau distribution. To improve the resolution of the \textit{dE/dx}
measurement, multiple samplings in pad rows along the track are performed.  An appropriate estimate of the \textit{dE/dx} 
is then calculated as a truncated mean of the distribution of deposited charge measurements. To obtain the contributions
of different particle species, fits of the inclusive \textit{dE/dx} distributions (see Ref.~\cite{dedx_Marco} for details)
were performed separately for negatively and positively charged particles in bins of total laboratory momentum $p$, 
transverse momentum ($p_{\bot}$) and azimuthal angle ($\phi$). Bins with less than 3000 entries were not 
used in the analysis to ensure sufficient statistics in each bin for the fitting algorithm. 
The distribution of the number of measured \textit{dE/dx} points in a representative bin is illustrated in 
Fig.~\ref{figPoints}. As for each track the energy loss is measured multiple times, the inclusive \textit{dE/dx} distribution 
(averaged over all events for the particular bin) for each particle type $j$ ($j$ = p, K, $\pi$, e) is represented by a weighted sum of Gaussian functions: 
 
 \begin{figure}[htbp]
\begin{center}
\includegraphics[width=0.5\textwidth]{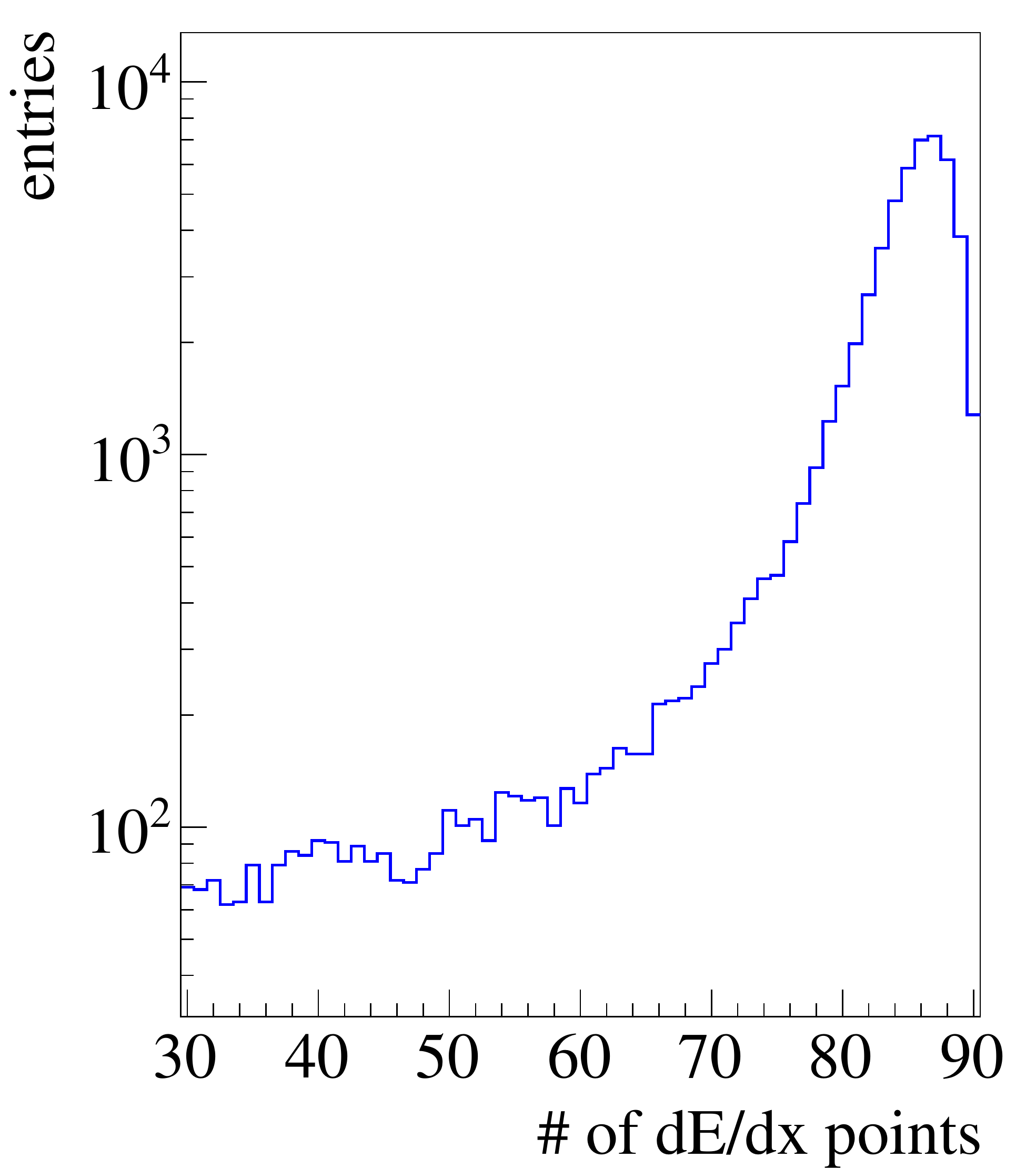}
\end{center}
\caption{(Color Online) Distribution of number of measured \textit{dE/dx} points along the tracks for the phase space bin
5.2 $ < p $ [GeV/c] $ < $ 6.4,  0.4 $< p_{\bot}$ [GeV/c]  $<$ 0.6  and  135 $ < \phi$ [$^{\text{o}}$]  $<$ 180 at 20\textit{A} GeV.}
\label{figPoints}
\end{figure}

\begin{figure}[htbp]
\begin{center}
\includegraphics[width=0.6\textwidth]{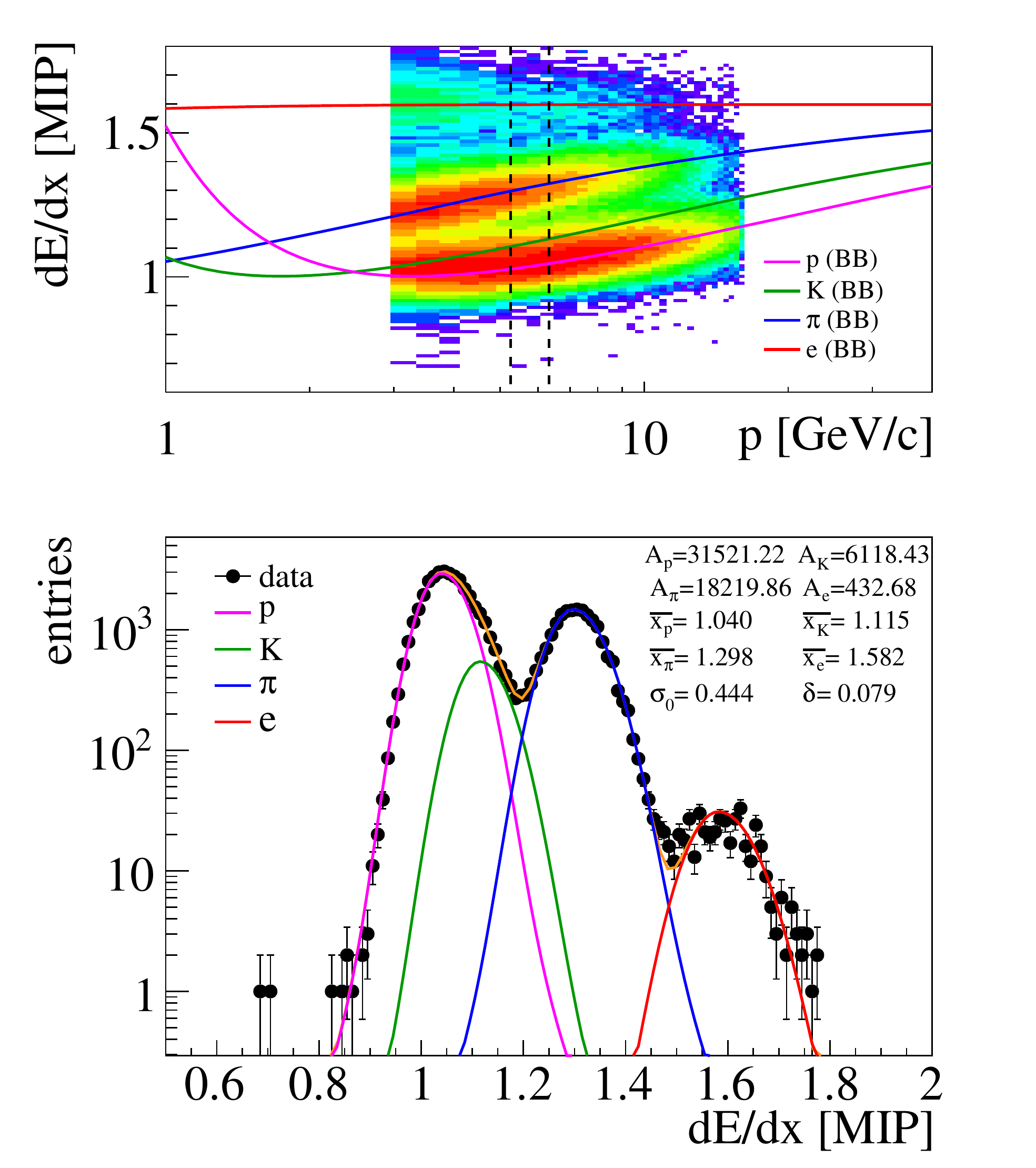}
\end{center}
\caption{(Color Online) Upper panel: Measured \textit{dE/dx} values as function of reconstructed momenta at 20\textit{A} GeV for the phase space region
 0.4 $< p_{\bot}$ [GeV/c] $<$ 0.6 and 135 $ < \phi$ [$^{\text{o}}$]  $<$ 180. Lines correspond to calculations with the Bethe-Bloch (BB) formula for
different particle types. Lower panel:  Projection of the upper plot  to the vertical axis in the momentum interval 
 5.2 $ < p$ [GeV/c] $<$ 6.4 indicated by vertical dashed lines.  Colored lines represent the \textit{dE/dx} distribution functions of different particles using Eq.~(\ref{dedx1}) and the fit parameters listed in the figure.
}
\label{figFits}
\end{figure}

 \begin{equation}
F_{j}\left(\frac{dE}{dx}\equiv x \right) = \frac{1}C\sum_{n}\frac{N_{n}}{\sqrt{2\pi}\sigma_{j,n}}\exp{\left[-\frac{1}{2}\left(\frac{x-\overline{x_j}}{(1\pm\delta)\sigma_{j,n}}\right)^{2}\right]}.
\label{dedx1}
\end{equation}
Here, $N_{n}$ is the number of tracks with $n$ \textit{dE/dx} measurements, 
$\overline{x_j}$ is the fitted mean energy loss (later referred to as position) of particle type $j$, 
and $\sigma_{j,n}$ is the width of the Gaussian distribution which depends on particle type $j$ and the number of \textit{dE/dx} measurements, $n$. The asymmetry parameter $\delta$ was introduced to account for the tails of the Landau distributions, which are still present even after truncation.
The normalization constant $C$ in Eq.(\ref{dedx1}) is $\sum_{n}N_{n}$, while $\sigma_{i,n}$ is parametrized as:

 \begin{equation}
\sigma_{j,n} = \sigma_{0}\left(\frac{\overline{x_{j}}}{\overline{x_{\pi}}}\right)^{\alpha}\frac{1}{\sqrt{N_{n}}},
\label{dedx2}
\end{equation}
where $\alpha$ was estimated from the data and set to 0.625~\cite{dedx_Marco}.

The parameterization of the total energy loss distribution is obtained by summing the functions $F_{j}$ over 
the particle types:

\begin{equation}
F(x) = \sum_{j=p,K,\pi,e}A_{j}F_{j}(x)
\end{equation}
with $A_{j}$ being the yield of particle $j$ in a given bin. 
As a result of fitting this function to the experimental \textit{dE/dx} distributions
one obtains in each phase space bin the yield of particle $j$, $A_{j}$, the ratio of 
mean ionization loss $\overline{x_{j}}/\overline{x_{\pi}}$, the parameter $\sigma_{0}$, and the asymmetry parameter $\delta$. 
The total number of fitted parameters is 2($k$+1) with $k$ denoting the number of particles. Obtained fit parameters, 
which are later used to access the \textit{dE/dx} distribution functions (DFs) of different particles, are stored in a lookup table.  
In the case of positive particles, DFs of kaons are masked by the protons and the mean values for protons and kaons cannot 
be fitted uniquely. To circumvent this problem the fitting procedure was performed in two steps:

\begin{enumerate}
\item The fitting procedure is started with negatively charged particles. As for the studied energy range the number of 
antiprotons is small, the pion and kaon peaks are essentially separated.
Furthermore, to enhance the statistics, integration is performed over the transverse momentum bins at this stage.
\item The fitting procedure is repeated separately for negatively and positively charged particles 
in bins of $p$, $p_{\bot}$ and $\phi$ with the ratio $\overline{x_{K}}/\overline{x_{\pi}}$ fixed from step 1. 
\end{enumerate}

As an example, we present in the upper panel of Fig.~\ref{figFits} a plot of measured \textit{dE/dx} values versus the 
reconstructed momenta. The lower panel of Fig.~\ref{figFits} shows the projection of the upper plot onto the \textit{dE/dx} axis 
in the selected momentum interval indicated by dashed vertical lines. The distribution functions of different particles 
obtained from Eq.(\ref{dedx1}) using the fit parameters listed in the figure are displayed by colored lines.

\begin{figure}[htbp]
\begin{center}
\includegraphics[width=0.6\textwidth]{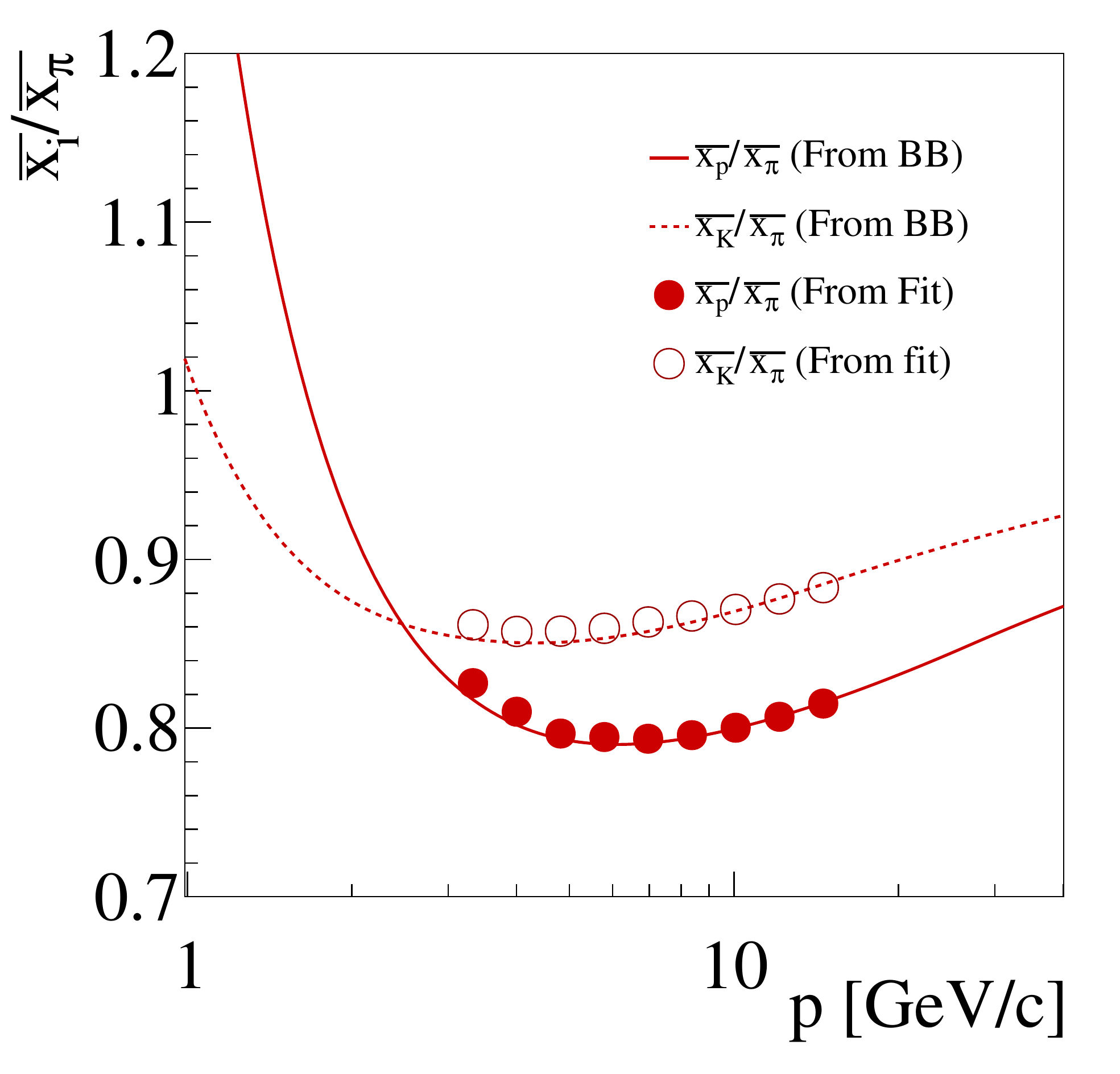}
\end{center}
\caption{(Color Online) Ratio of fitted mean energy losses (symbols) compared to corresponding ratios 
from the Bethe-Bloch parametrization (curves) for 20$A$ GeV data. The deviations of the fitted values from the Bethe-Bloch curves are below 1 $\%$. 
}
\label{figQuality}
\end{figure}

\begin{figure}[htbp]
\begin{center}
\includegraphics[width=0.6\textwidth]{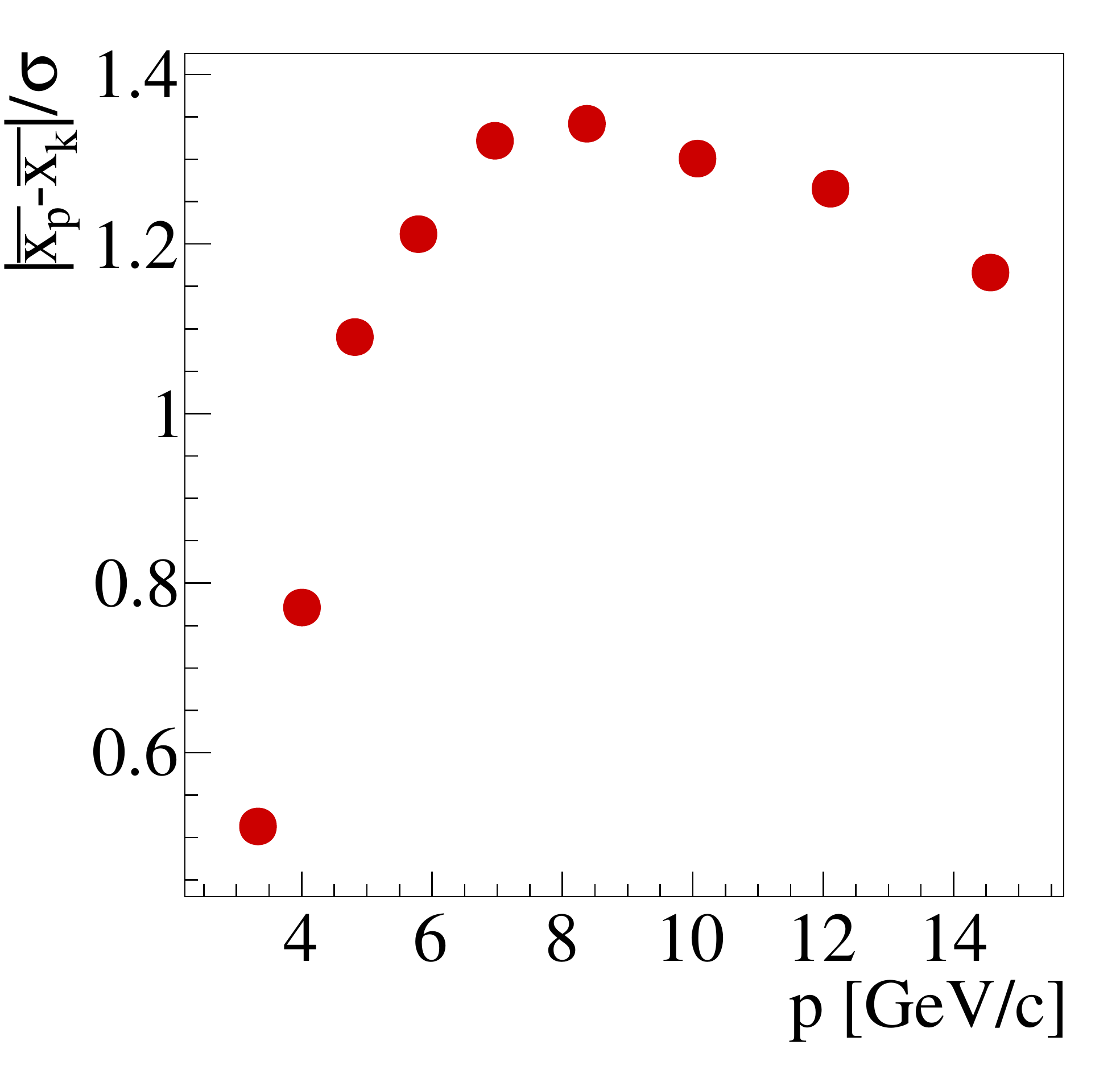}
\end{center}
\caption{(Color Online) The difference between mean energy loss of kaons and protons normalized to the \textit{dE/dx} width for 20\textit{A} GeV data.
}
\label{separation}
\end{figure}

In Fig.~\ref{figQuality} the ratios of mean energy losses  of different particles are compared to the corresponding 
ratios from the Bethe-Bloch parameterization. Figure~\ref{separation} demonstrates the separation between fitted mean 
energy loss values of kaons and protons quantified as  $\left| \overline{x_{p}}-\overline{x_{K}}\right | /  \sigma$ with $x_{p}$ and $x_{K}$ denoting the 
mean energy loss values for protons and kaons respectively, and $\sigma$ stands for $\sqrt{\sigma_p^{2} + \sigma_K^{2}}$. Here the $\sigma_{j}$ ($j$ = p, K)
is calculated as:

\begin{equation}
\sigma_{j}=\frac{1}{C}\sum_{n}\sigma_{j,n},
\end{equation}
with $C$ and $\sigma_{j,n}$ defined in Eqs.~(\ref{dedx1}) and~(\ref{dedx2}).

 \section{Analysis Method}
 \label{sec-identity}
Most measures proposed for event-by-event fluctuations are defined as functions of moments of the unknown 
multiplicity distributions. In particular, the fluctuation measure $\nu_{\text{dyn}}$ depends on the first and all second 
(pure and mixed) moments of the multiplicity distributions of the studied particles species. For example, second (pure) moment for pions and the second mixed moment for protons and pions are defined as:

\begin{equation}
\langle N_{\pi}^{2}\rangle = \sum_{N_{\pi}=0}^{\infty}N_{\pi}^2P(N_{\pi}),
\label{defMoments1}
\end{equation}

and

\begin{equation}
\langle N_{\pi}N_{p}\rangle = \sum_{N_{\pi}=0}^{\infty}\sum_{N_{p}=0}^{\infty}N_{\pi}N_{p}P(N_{p},N_{\pi}),
\label{defMoments2}
\end{equation}
where, $P(N_{\pi})$ is the probability distribution of pion multiplicity, while $P(N_{p},N_{\pi})$ is the joint probability distribution for pion and proton multiplicities.
$N_{\pi}$ and $N_{p}$ in Eqs.~(\ref{defMoments1}) and~(\ref{defMoments2}) stand for the pion and proton multiplicities. 

The standard approach of finding the moments is to count the number of particles event-by-event.  
However, this approach is hampered by incomplete particle identification (overlapping \textit{dE/dx} distribution functions), 
which can be taken care of by either selecting suitable phase space regions (where the distribution functions do 
not overlap) or by applying an event-by-event fitting procedure . The latter typically introduces artificial 
correlations which are usually corrected for by the event mixing technique. Here a novel approach, 
called \emph{Identity Method}~\cite{identity1,identity2,identity3}, is applied for the first time.
The method follows a probabilistic approach which avoids the event-by-event fitting and
determines the moments of the multiplicity distributions by an unfolding procedure which 
has a rigorous mathematical derivation~\cite{identity2}. Thus there is no need for corrections based on event mixing.
The method employs the fitted inclusive \textit{dE/dx} distribution functions of particles, 
$\rho_{j}(x)$, with $j$ standing for proton, kaon, pion and electron. Each event has a set of 
measured \textit{dE/dx} values, $x_{i}$, corresponding to each track in the event. For each track in an event 
a probability $w_j$ was estimated of being a particle $j$:

\begin{figure}[htbp]
\begin{center}
\includegraphics[width=0.7\textwidth]{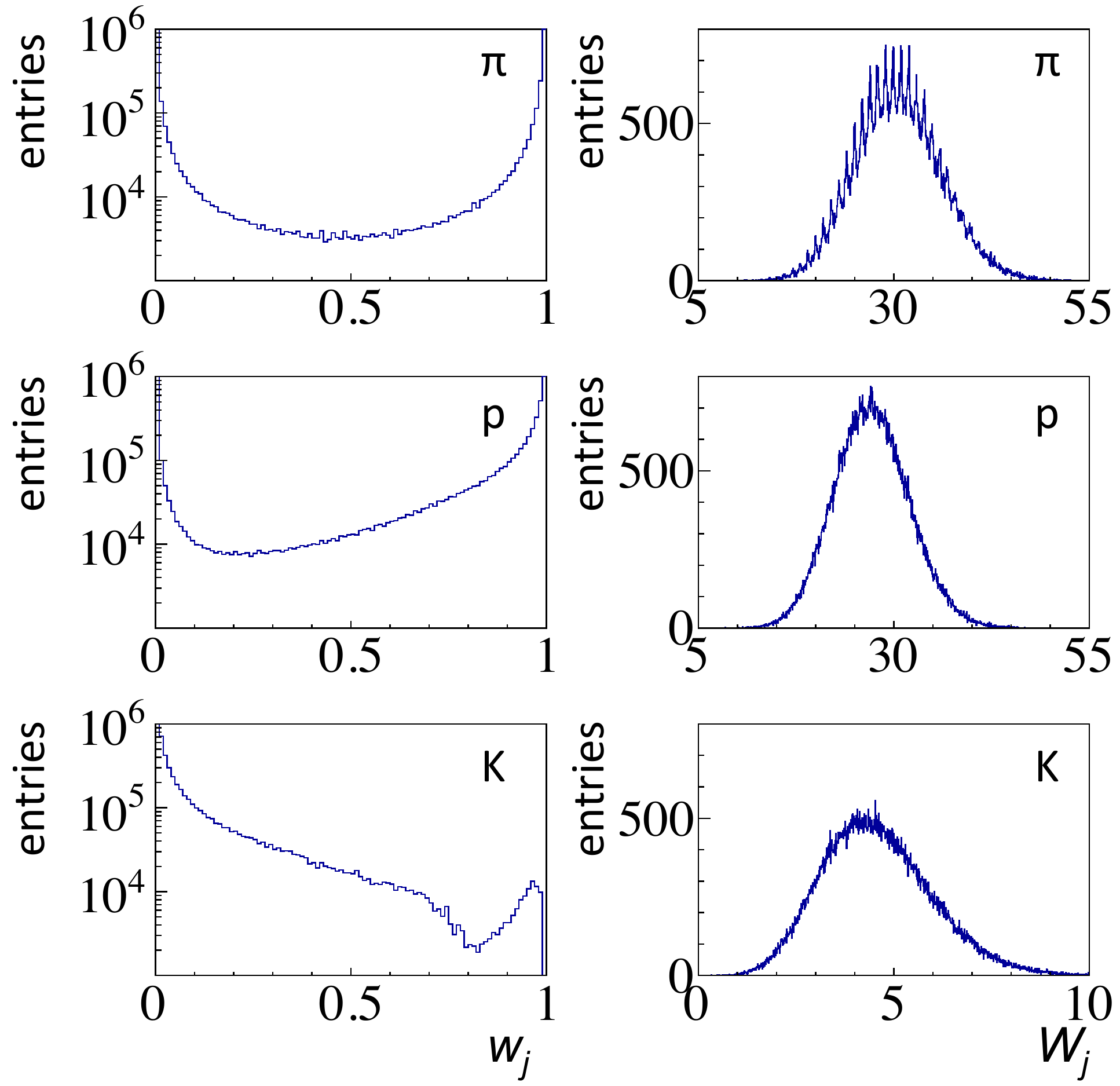}
\end{center}
\caption{(Color Online) Distributions of $w_j$ of Eq.(\ref{eq_w}) and $W_j$ of Eq.(\ref{eq_W}) 
for different particle types $j$ for 20\textit{A} GeV data. 
}
\label{w_dist}
\end{figure}

 \begin{center}
 \begin{table*}
\begin{tabular}{|c|c|c|c|c|c|c|}
\hline 
 & 20\textit{A} GeV & 30\textit{A} GeV & 40\textit{A} GeV & 80\textit{A} GeV & 160\textit{A} GeV\\ [0.5ex] 
\hline  
$\left< N_p \right>$ & 27.1 & 34.7 & 38.0 & 47.0 & 68.7 \\  [0.5ex] 
\hline  
$\left<N_\pi\right>$ & 30.5 & 66.4 & 103.0 & 226.7 & 414.6  \\  [0.5ex] 
\hline  
 $\left<N_K\right>$& 4.7 & 9.4 & 13.9 & 31.5 & 57.8 \\  [0.5ex] 
\hline  
\hline
\hline
$\left<N_p^2 \right>$& 759.94 & 1238.09 &1475.89 & 2254.35 & 4780.52 \\  [0.5ex] 
\hline  
$\left<N_\pi^2\right>$& 963.6 & 4485.36 & 10731.4 & 51764.4 & 172811.0 \\  [0.5ex]
 \hline
 $\left<N_K^2\right>$& 26.4 & 98.06 & 207.27 & 1030.06 & 3415.69 \\  [0.5ex]
 \hline 
 Cov$\left[N_{p},N_{\pi}\right]$ & 2.13 & 4.34 & 9.05 & 22.62 & 44.03 \\  [0.5ex]
 \hline
 Cov$\left[N_p,N_K\right]$& -0.75 & -0.69 & 0.39 & 2.41 & 10.92 \\  [0.5ex]
 \hline
 Cov$\left[N_K,N_{\pi}\right]$& -1.02 & -1.39 & 0.29 & 15.84 & 81.75 \\  [0.5ex]
 \hline
 \end{tabular}
\caption{Upper part: mean multiplicities of $p+\bar{p}$, $\pi^{+}+\pi^{-}$, and $K^{+}+K^{-}$ 
for the 3.5$\%$ most central Pb+Pb collisions calculated by summing the integrals of respective DFs 
over phase-space bins. Lower part: reconstructed second moments of the multiplicity distributions 
of $p+\bar{p}$, $\pi^{+}+\pi^{-}$, and $K^{+}+K^{-}$ for the 3.5$\%$ most central Pb+Pb collisions. 
The mixed moments are presented in terms of covariances, Cov$[N_{1},N_{2}]=\langle N_{1}N_{2} \rangle -\langle N_{1} \rangle \langle N_{2} \rangle$. 
For 20\textit{A} and 30\textit{A} GeV, values for Cov$[N_{p},N_{K}]$ and Cov$[N_{p},N_{K}]$ are negative.
Numerical values with higher precision are available in Ref~\cite{EDMS}. These are required to reproduce the 
values of $\nu_{\text{dyn}}$ shown in this paper.
}
\label{tableMoments}
\end{table*}
\end{center}

\begin{equation}
w_{j}(x_{i}) \equiv \frac{\rho_{j}(x_{i})}{\rho(x_{i})},
\label{eq_w}
\end{equation}
where the values of $\rho_{j}(x_{i}) = A_{j}F_{j}(x_{i})$ are calculated using the parameters stored 
in the lookup table of fitted DFs in the appropriate phase space bin, and 

\begin{equation}
\rho(x_{i}) \equiv \sum_{j=p,K,\pi,e}\rho_{j}(x_{i}).
\end{equation}

Note that the $\rho_j$ functions are just DFs normalized to the total number of events.
Further an event variable (an approximation of the multiplicity of particle $j$ in the event) $W_{j}$ is defined as:

\begin{equation}
W_{j}=\sum_{i=1}^{n}w_{j}(x_{i}),
\label{eq_W}
\end{equation}
where $n$ is the total number of selected tracks in the given event. Examples of distributions of $w_{j}$ and $W_{j}$
for $\pi$, $K$ and $p$ are shown in Fig.~\ref{w_dist}. 

As the introduced $W_{j}$ quantities are calculated for each event, one obtains all second moments of the $W_{j}$ quantities
by straightforward averaging over the events. Finally, using the \textit{Identity Method} one unfolds the second moments 
of the true multiplicity distributions from the moments of the $W_{j}$ quantities~\cite{identity2}.
Obtained results (second moments) for the $3.5\%$ most central Pb+Pb collisions at different projectile energy
are listed in the lower part of Table~\ref{tableMoments}. The mean multiplicities (first moments) shown 
in the upper part of Table~\ref{tableMoments} are the results of integration of the respective DFs.
The Identity Method has been successfully tested for numerous simulations in Ref.~\cite{identity3}. 
A direct experimental verification of the method can be provided by investigating the energy dependence 
of the scaled variance $\omega$ of the negatively charged pion multiplicity distribution, where $\omega$ is
\begin{equation}
\omega=\frac{\text{Var}(N)}{\langle N \rangle} = \frac{\langle N^2 \rangle - \langle N \rangle ^2}{\langle N \rangle}.
\label{omega_eq_1}
\end{equation}

\begin{figure}[htbp]
\begin{center}
\includegraphics[width=0.6\textwidth]{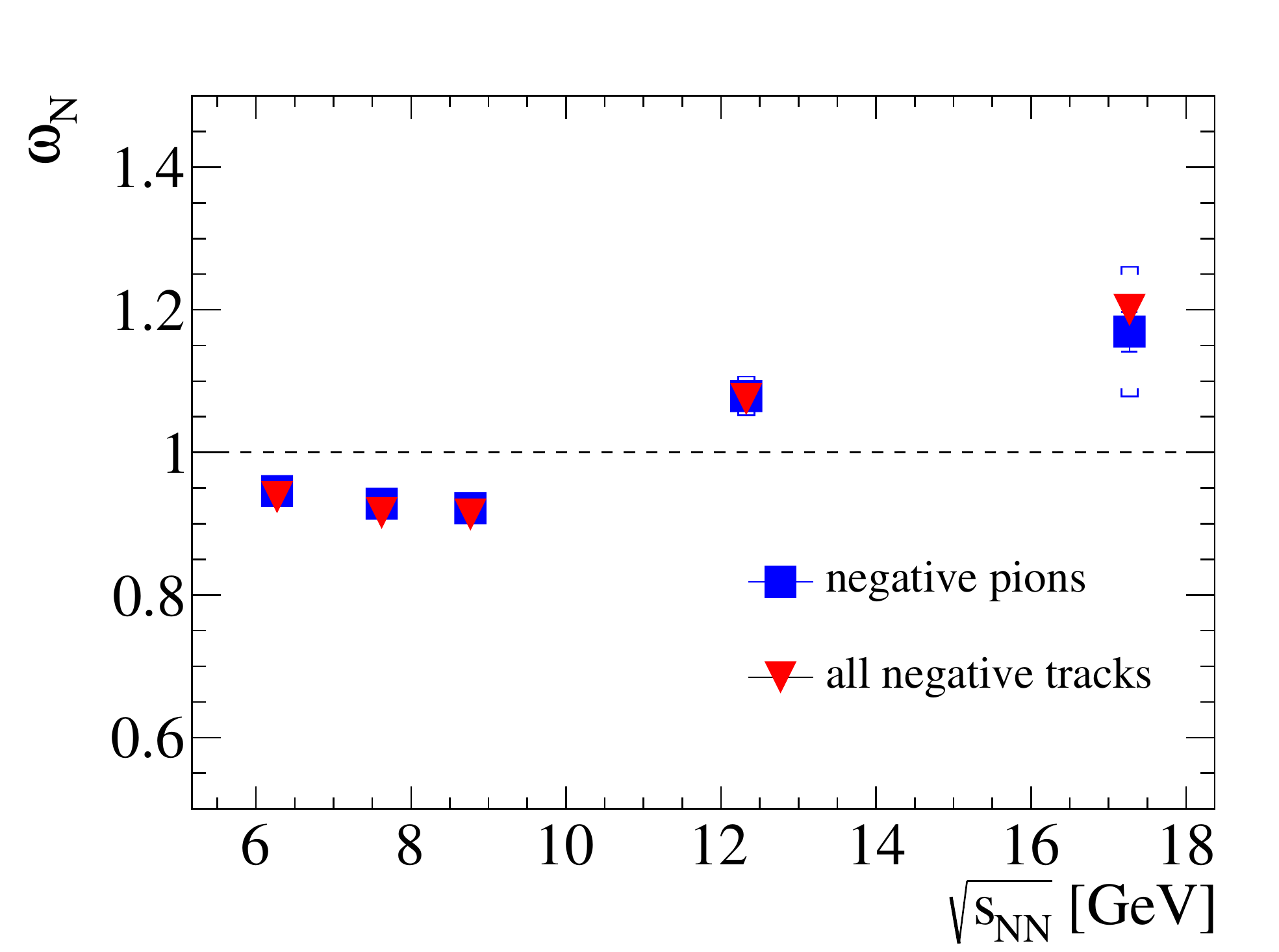}
\end{center}
\caption{(Color Online) The energy dependence of the scaled variance $\omega$ of the negatively charged pion
multiplicity distribution, reconstructed using the \textit{Identity Method}, 
is plotted as blue squares. The red triangles are estimates based  on direct 
event-by-event counting of all negative particles. The remarkable agreement between these results 
is an experimental verification of the \textit{Identity Method}. 
}
\label{omega2}
\end{figure}	

For this purpose two independent analyses were performed: (i) using the reconstructed moments for negatively 
charged pions (from the \textit{Identity Method}) and (ii) counting the negatively charged particles event-by-event 
(i.e., without employing the Identity Method). The results of these analyses are presented in Fig.~\ref{omega2} 
by blue squares for case (i) and by red triangles for case (ii). As the majority of negative particles 
are pions the remarkable agreement between the results of these two independent approaches is a direct 
experimental verification of the \textit{Identity Method}.

\section{Statistical and systematic error estimates}
\label{sec-errors}

\begin{figure}[htbp]
\begin{center}
\includegraphics[width=0.5\textwidth]{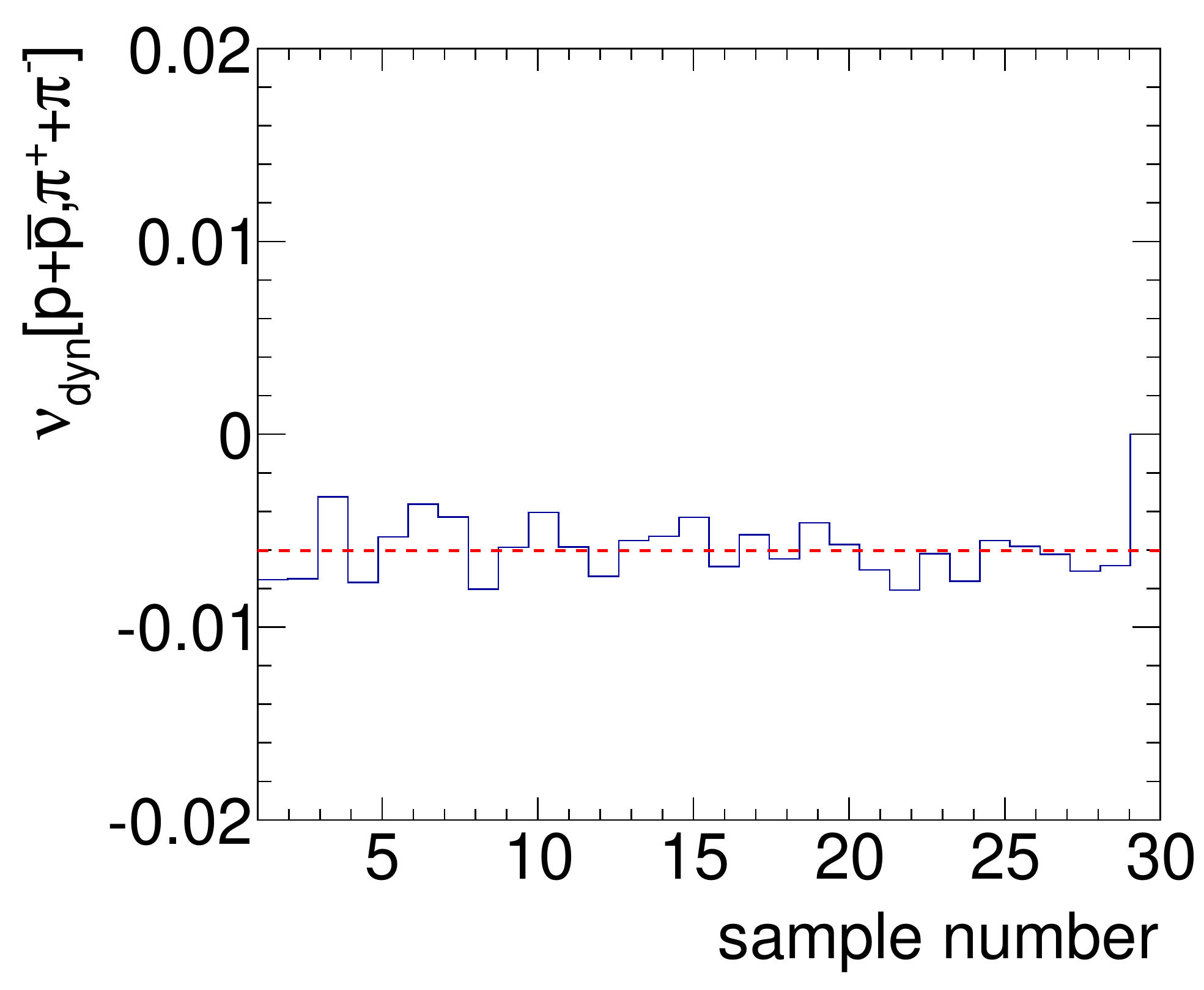}
\end{center}
\caption{(Color Online) Reconstructed values $\nu_{\text{dyn}}[p+\bar{p},\pi^{+}+\pi^{-}]$ as a function of subsample number. The dashed red line indicates the averaged value of $\nu_{\text{dyn}}$ over subsamples. 
}
\label{nu_dyn_subsamples}
\end{figure}

The statistical errors of the reconstructed moments of the
multiplicity distributions result from the errors on the
parameters of the fitted distributions $\rho_{j}(x)$  and from
the errors of the $W_{j}$ quantities. Typically these two sources
of errors are correlated. Fluctuation observables are usually built up from several moments of
the multiplicity distributions. Since the standard error propagation is impractical,
the subsample approach was chosen to evaluate the statistical
uncertainties. One first randomly subdivides the data into $n$
subsamples and for each subsample then reconstructs the moments $M_{n}$ 
listed in Table~\ref{tableMoments}. In the second step the statistical error of each moment $M$ 
is calculated as:
\begin{equation}
 \sigma_{\langle M \rangle}=\frac{\sigma}{\sqrt{n}},
\label{stat-1}
\end{equation}
where
\begin{equation}
 \langle M \rangle =\frac{1}{n}\sum{M_n} ,
\label{stat-2}
\end{equation}
and
\begin{equation}
 \sigma=\sqrt{\frac{\sum\left(M_i - \langle M \rangle \right)^2}{n-1}}.
\label{stat-3}
\end{equation}
The same procedure is followed for the fluctuation quantities, e.g., $\nu_{\text{dyn}}$, which are functions of the
moments. An example is shown in Fig.~\ref{nu_dyn_subsamples}.

Next, systematic uncertainties of the analysis procedure are discussed.
One possible source of systematic bias might be the specific choice of event and track cuts. In order to obtain
an estimate of this uncertainty, results for the moments were derived for "loose" and "tight" cuts (see scetion~\ref{sec-cuts}). 
The small observed differences were taken as one component of the systematic error.
	
\begin{figure}[htbp]
\begin{center}
\includegraphics[width=0.62\textwidth]{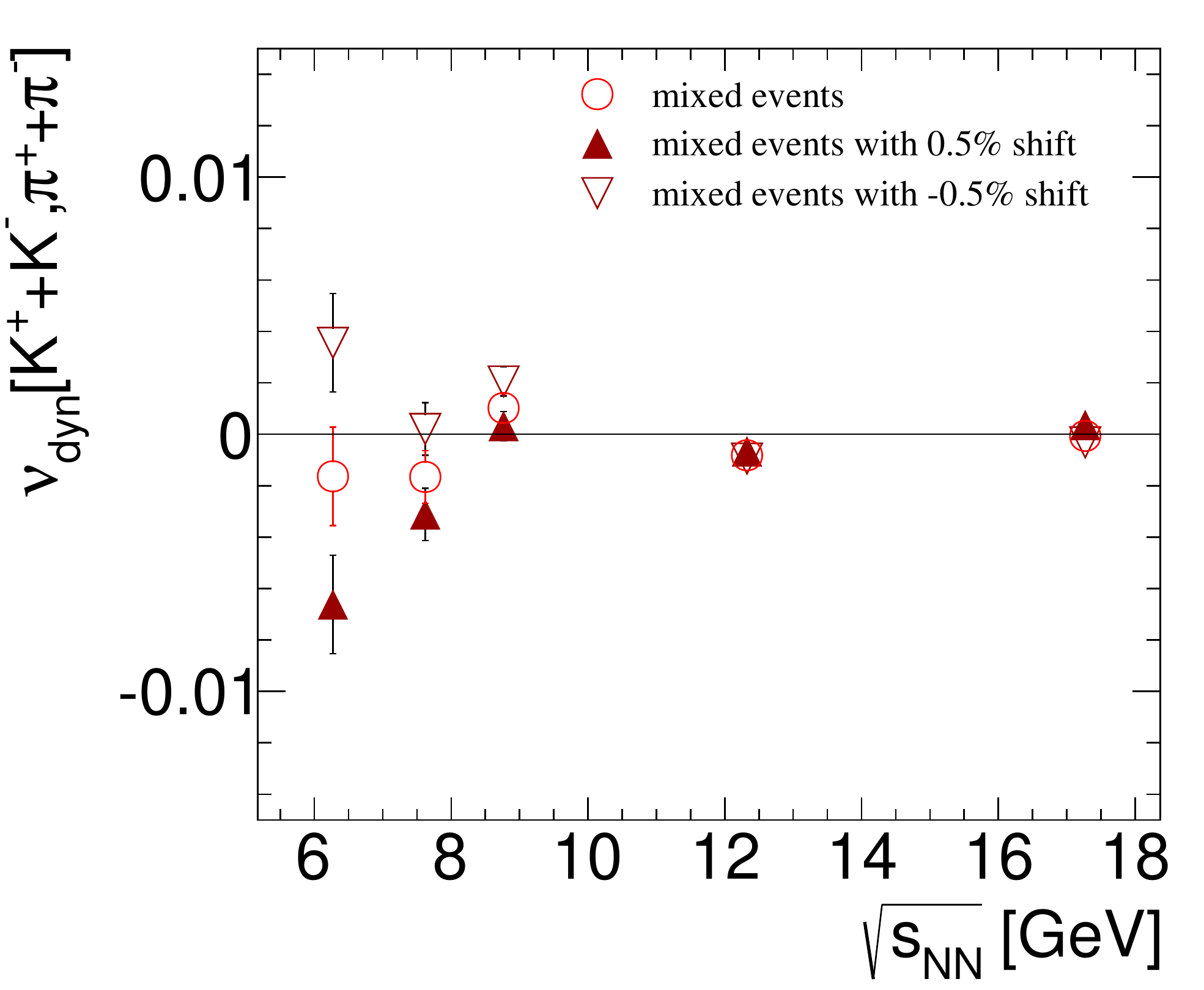}
\end{center}
\caption{(Color Online) $\nu_{\text{dyn}}[K^{+}+K^{-},\pi^{+}+\pi^{-}]$ for mixed events is shown versus energy by red open circles. 
Solid (open) red triangles represent the results obtained with the kaon positions shifted artificially by 0.5$\%$ (-0.5$\%$).
}
\label{figMixed}
\end{figure}

Possible biases of the identification procedure were studied using mixed events. Each event $i$ was constructed by randomly selecting
a reconstructed track (including the \textit{dE/dx} measurement) from each of the following $j$ events, with $j$ corresponding to the number of
reconstructed tracks in the event $i$. The results for $\nu_{\text{dyn}}[K^{+}+K^{-},\pi^{+}+\pi^{-}]$ for mixed events are 
presented in Fig.~\ref{figMixed} by red open circles. As expected the reconstructed values of $\nu_{\text{dyn}}$ are vanishing 
independently of energy.  

\begin{figure}[htbp]
\begin{center}
\includegraphics[width=0.45\textwidth]{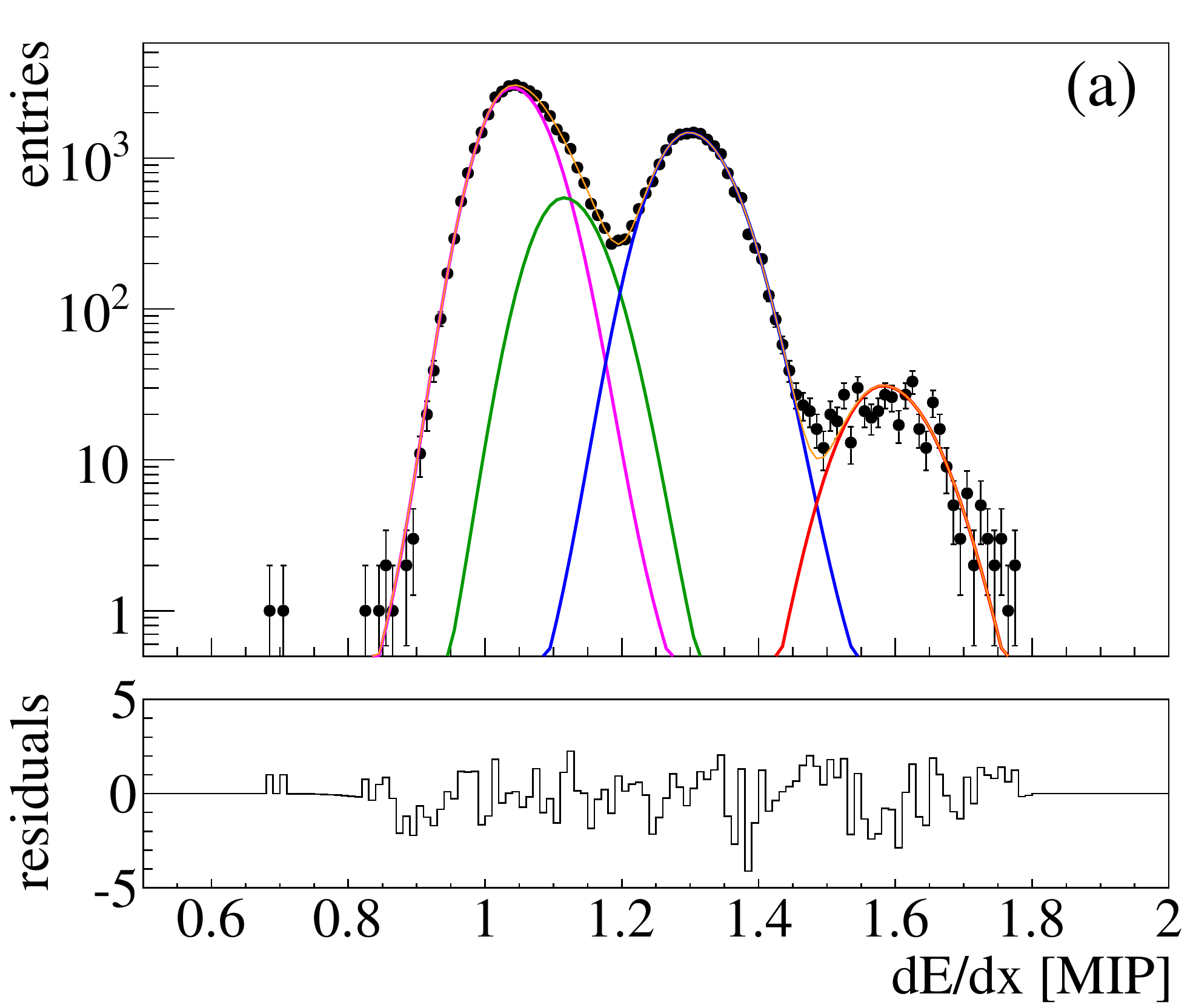}
\includegraphics[width=0.45\textwidth]{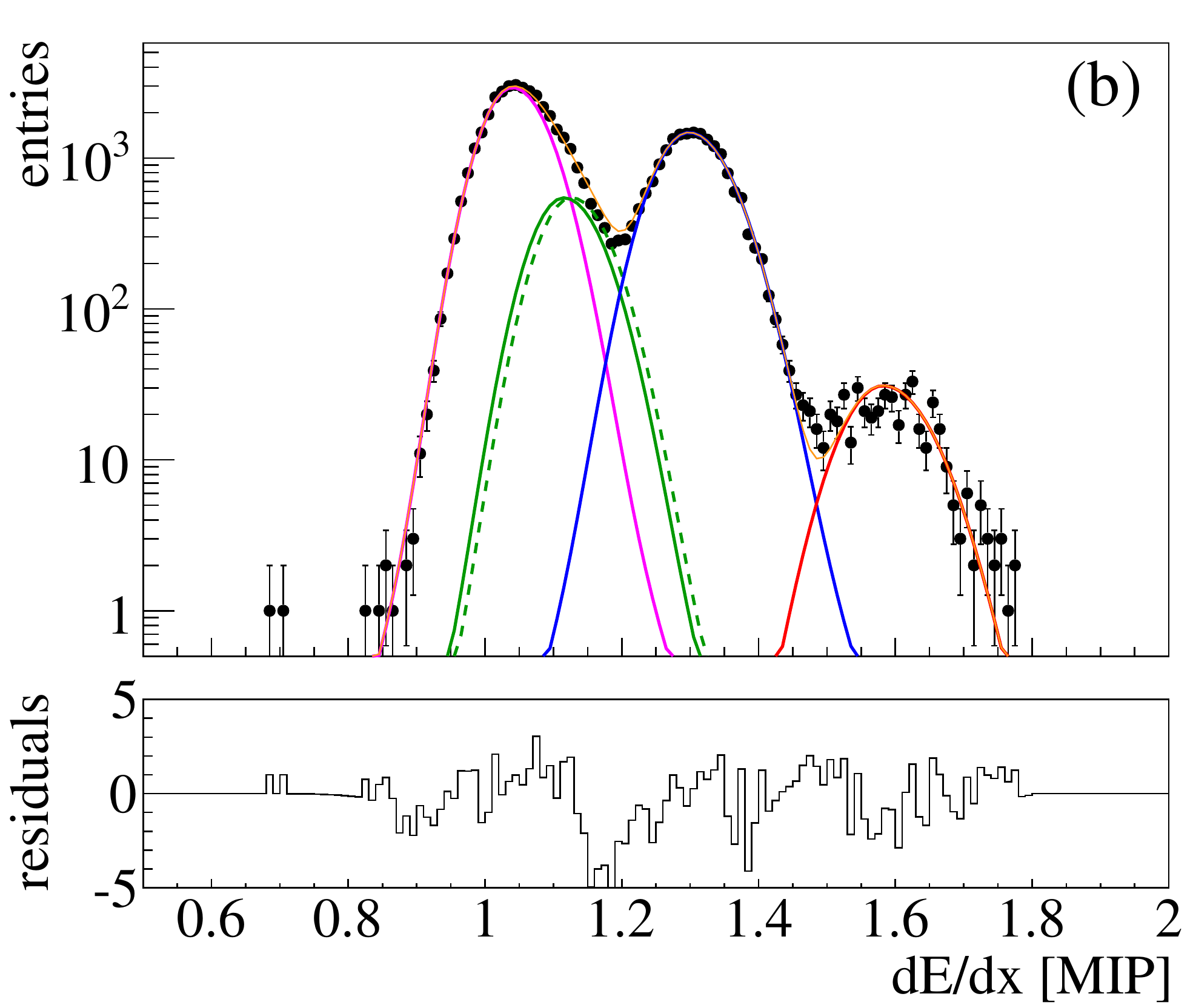}
\vspace{2mm}
\includegraphics[width=0.45\textwidth]{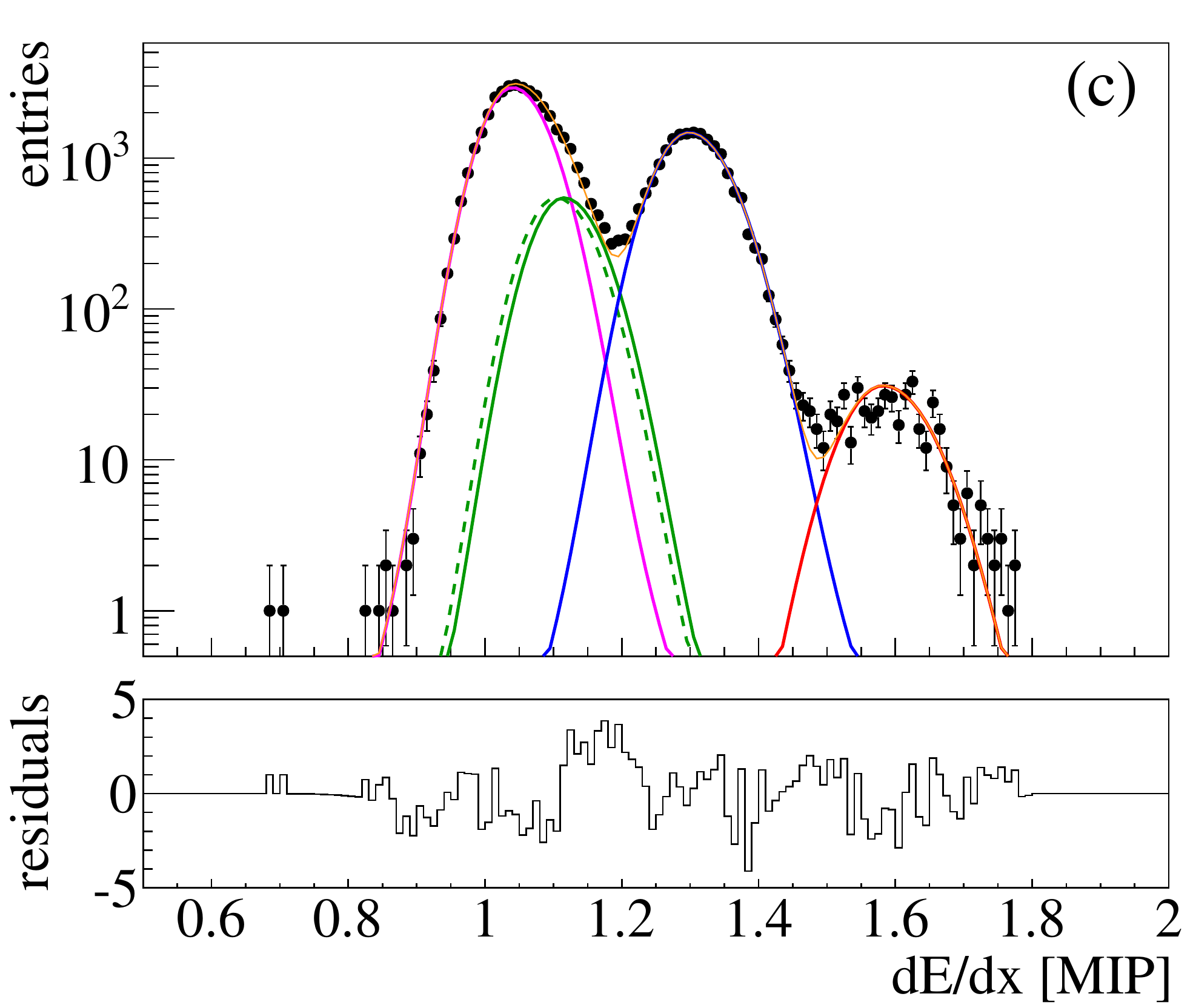}
\includegraphics[width=0.45\textwidth]{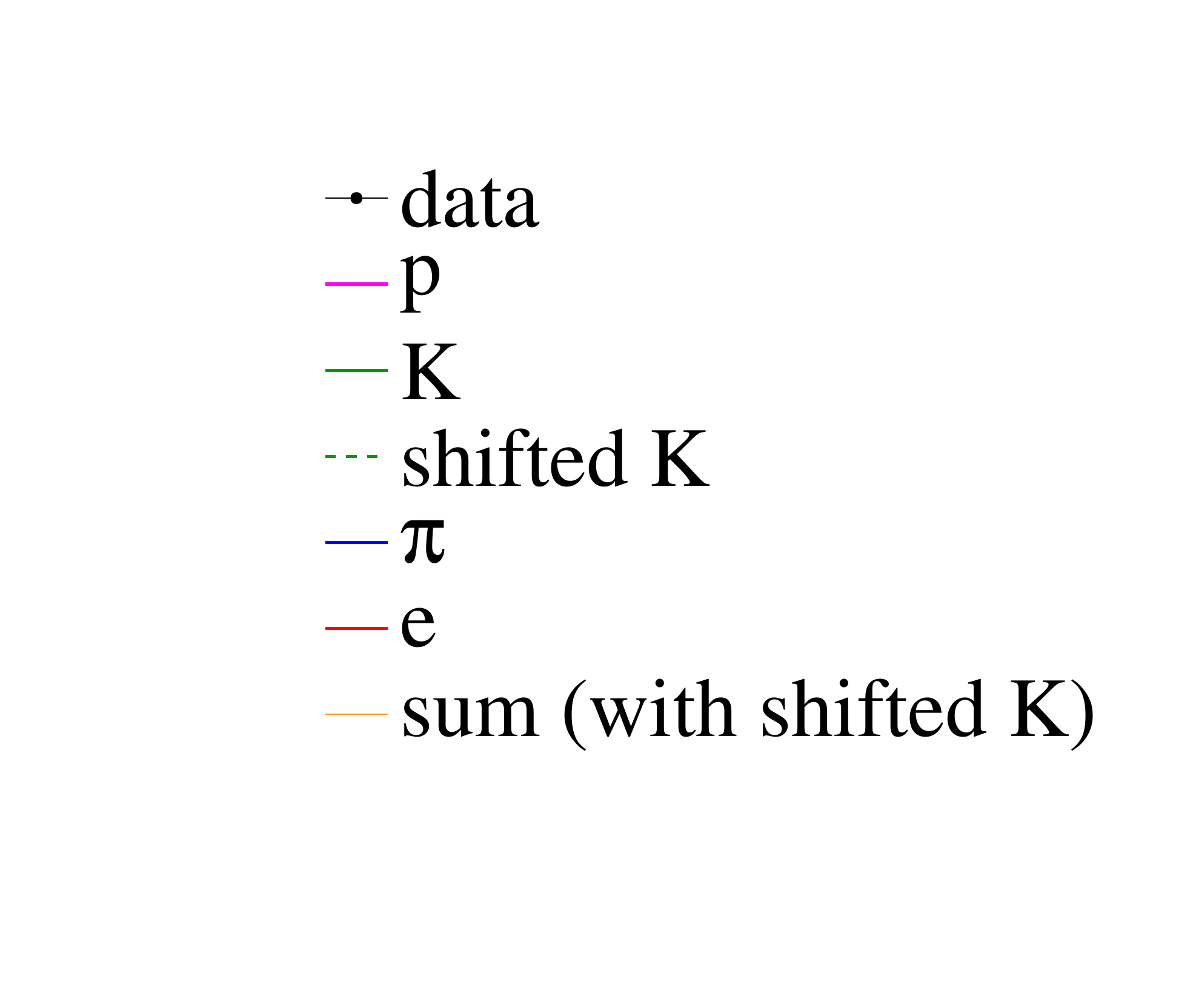}

\end{center}
\caption{(Color Online) Energy loss distributions in the selected phase space bin corresponding to Fig.~\ref{figFits} 
with superimposed fit functions for protons, pions, kaons and electrons shown by colored solid lines. 
The dashed green lines correspond to artificially shifted positions of kaons by 1$\%$ (b) and -1$\%$ (c). 
The shifted distribution functions were used to investigate the systematic errors stemming from the 
particle identification (\textit{dE/dx} fitting) procedure. The corresponding residual plots are also presented. The residuals are defined as the difference between data points and the total fit function (indicated by sum), 
normalized to the statistical error of data points.}
\label{shiftKaons}
\end{figure}

\begin{figure}[htbp]
\begin{center}
\includegraphics[width=0.45\textwidth]{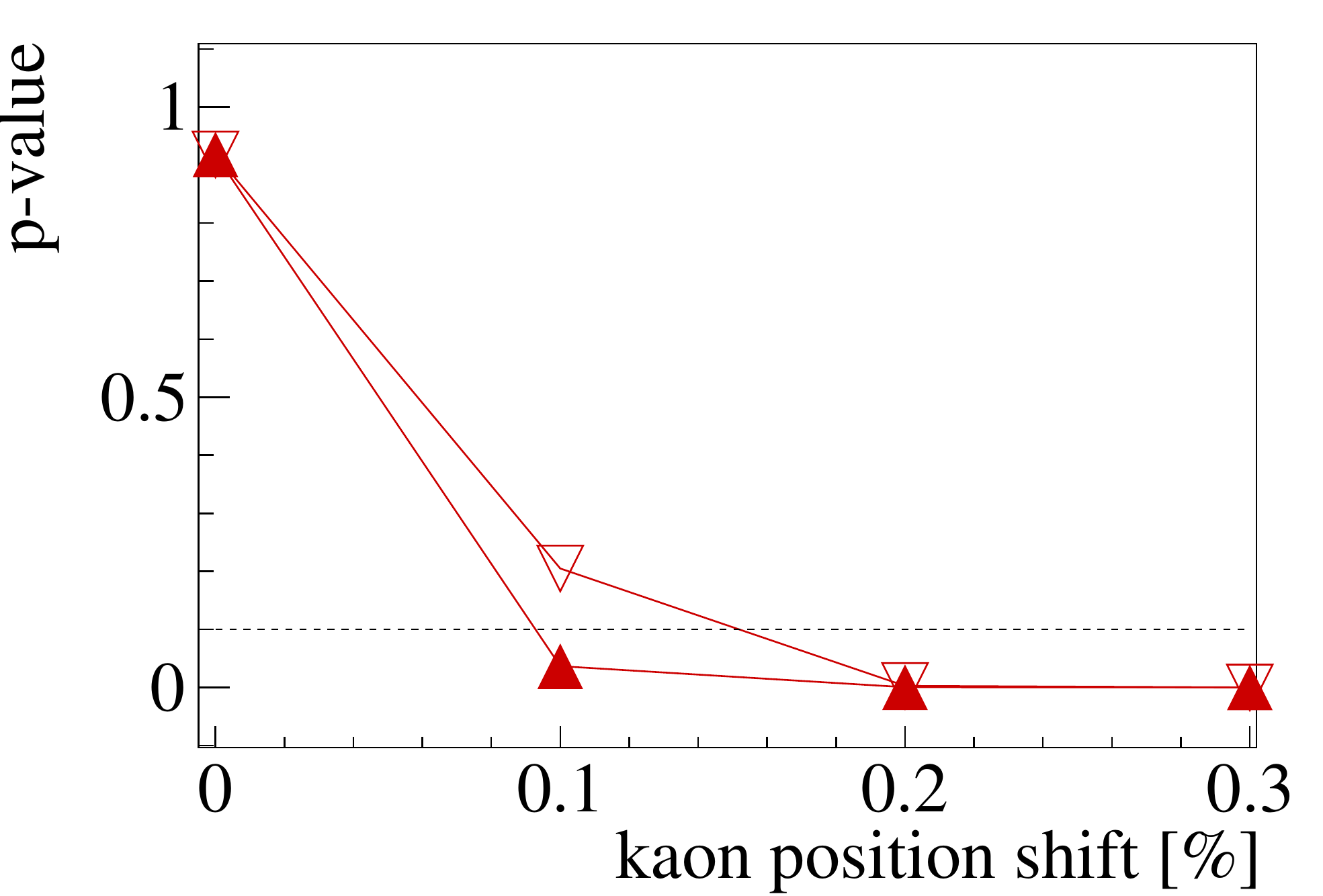}
\includegraphics[width=0.45\textwidth]{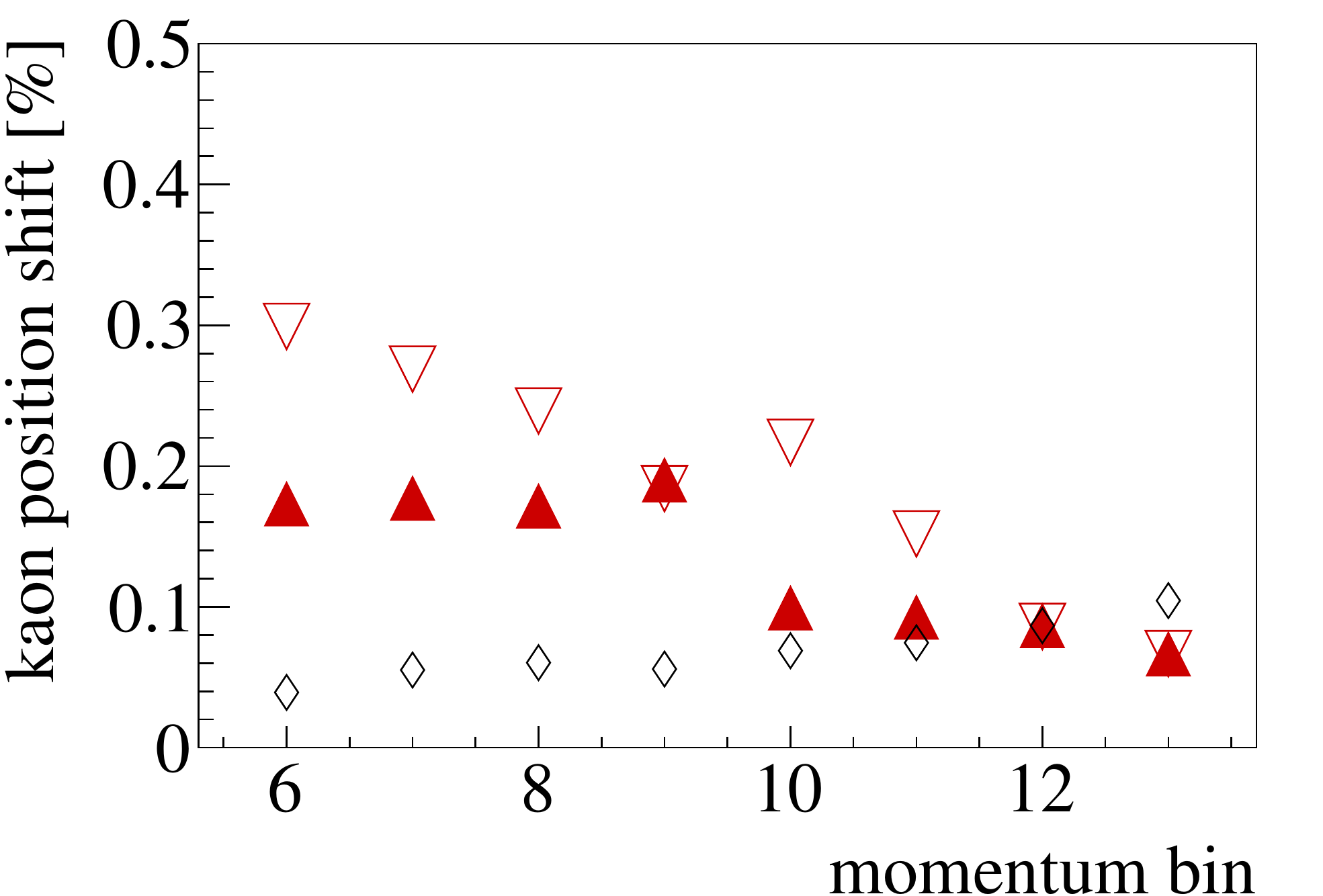}
\end{center}
\caption{(Color Online) Left panel: The p-value of the K-S statistics as function of the artificially introduced shifts 
in the fitted kaon positions for 30\textit{A} GeV data. The direction of triangles indicates the direction of  introduced shifts. The null 
hypothesis is rejected when the p-value is below the significance level of 10 $\%$, indicated by the dashed line. 
The maximum value of the kaon shift is taken as the abscissa of the intersection point of full red and dashed black lines. 
Right panel: Maximum values of the kaon position shift as function of the momentum in a selected bin of transverse 
momentum and azimuthal angle. Diamonds represent the statistical errors on kaon positions obtained from fitting procedure.
Note that the left plot corresponds to momentum bin 11.}
\label{K-S1}
\end{figure}

Furthermore, systematic uncertainties stemming from the quality of the fit functions were investigated 
with the help of mixed events. Even though the 2-step fitting procedure discussed in 
section~\ref{sec-pid} was used to determine the DFs, it remains a challenge to properly fit 
the kaon positions. In nearly all relevant phase-space intervals the measured energy loss 
distributions of kaons are overlapping with those of pions and protons. To study the influence of possible 
systematic shifts in fit parameters on the extracted moments, the fitted positions of kaons were
shifted artificially by 0.5 $\%$ in both directions. The dashed-green lines in Fig.~\ref{shiftKaons} 
show the artificially shifted \textit{dE/dx} distribution functions of kaons. 
Results for $\nu_{\text{dyn}}[K^{+}+K^{-},\pi^{+}+\pi^{-}]$ obtained with these shifted kaon distribution functions 
for the mixed events are plotted as red triangles in Fig.~\ref{figMixed}.
At lower beam energies one observes a significant dependence of the results on kaon positions. In order to gain quantitative estimates of
a possible shift of the kaon position, we performed hypothesis testing using the Kolmogorov-Smirnov (K-S) statistics.
For this purpose we test the null hypothesis that measured \textit{dE/dx} distributions and fit functions are similar within a given significance level of 10 $\%$. We repeat the test by shifting the fitted kaon positions in both directions. The obtained results from the K-S test in a selected phase space bin are presented in the left panel of Fig.~\ref{K-S1} for the 30\textit{A} GeV data. The maximum value of the kaon position shift is taken to be the abscissa of the intersection point of the red lines with the dashed line.
 We conclude that with a 10 $\%$ significance level the null hypothesis is rejected for 0.09 and 0.15 $\%$  up and down shifts correspondingly. 
In the right panel of Fig.~\ref{K-S1} the dependence of the kaon position shift is presented as function of the momentum bin 
in a selected bin of transverse momentum and azimuthal angle. 
The shift values for all other phase space bins were obtained in a similar way. 
Emerging systematic errors on the fluctuation measure $\nu_{\text{dyn}}$, added in quadrature 
with other sources of systematics,  are depicted in Fig.~\ref{nu_dyn} by the shaded bands (see the next section).

\section{Results on the fluctuation measure $\nu_{\text{dyn}}$}
\label{sec-nu}

\begin{figure}[htbp]
\begin{center}
\includegraphics[width=0.45\textwidth]{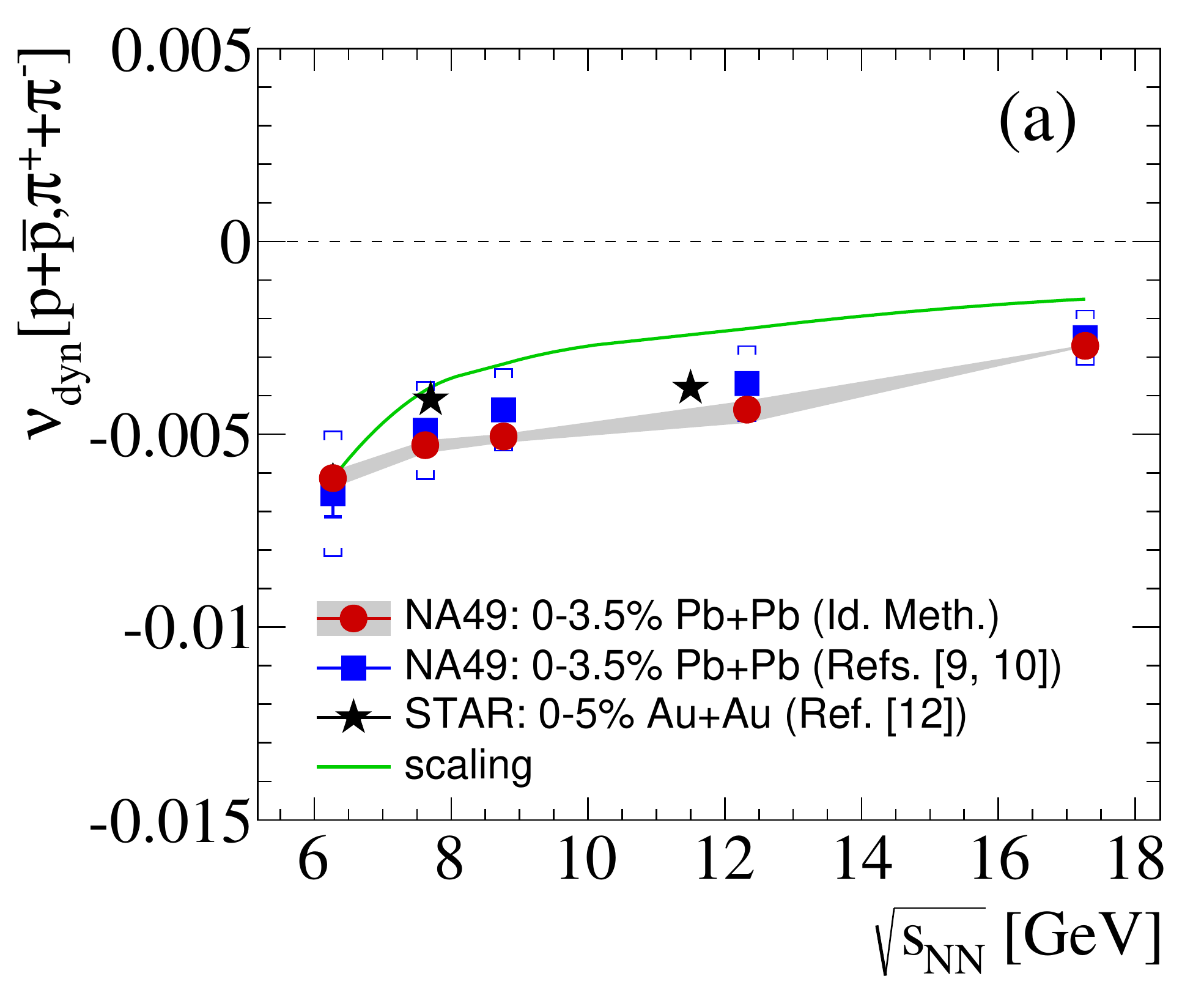}
\includegraphics[width=0.45\textwidth]{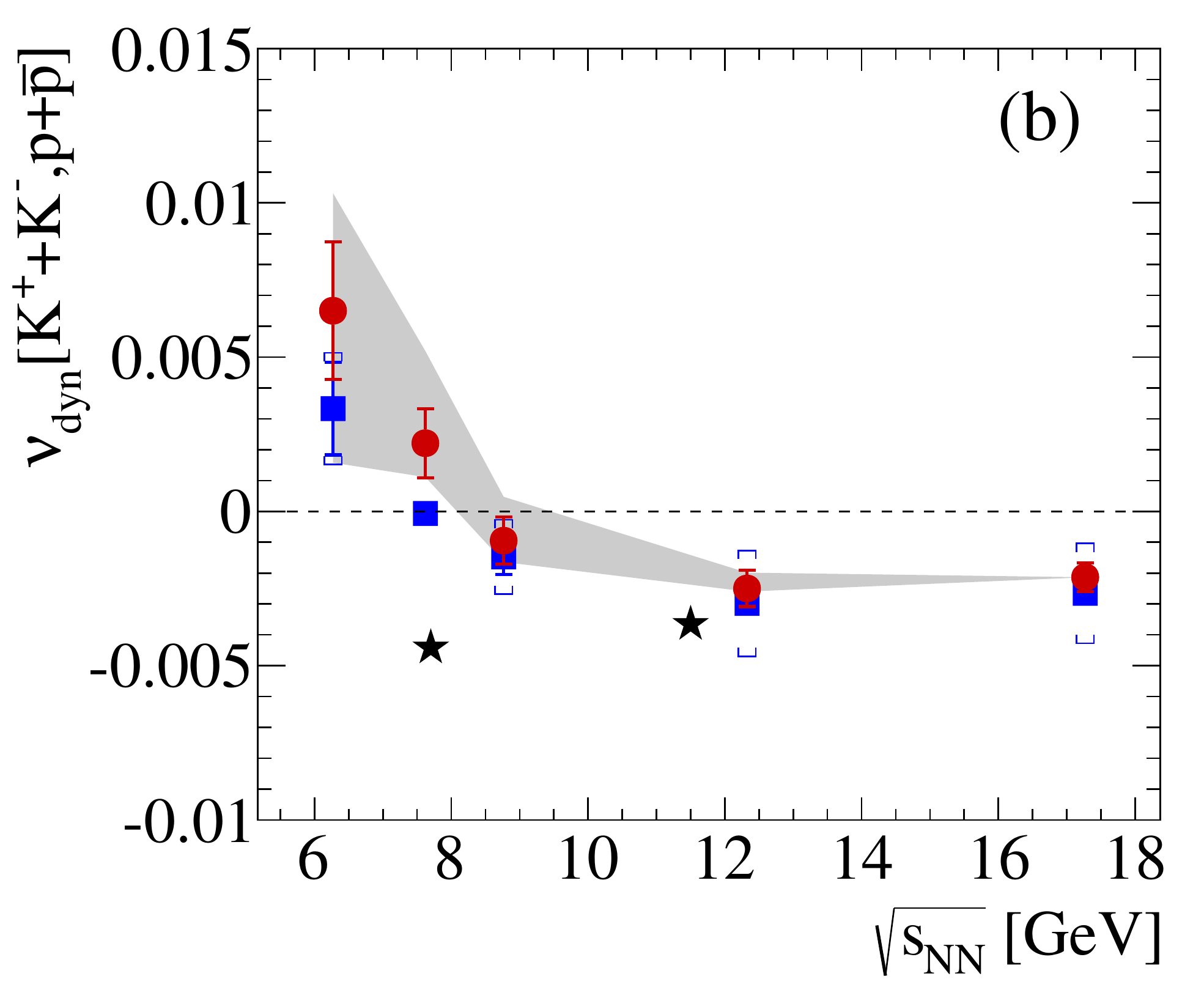}
\includegraphics[width=0.45\textwidth]{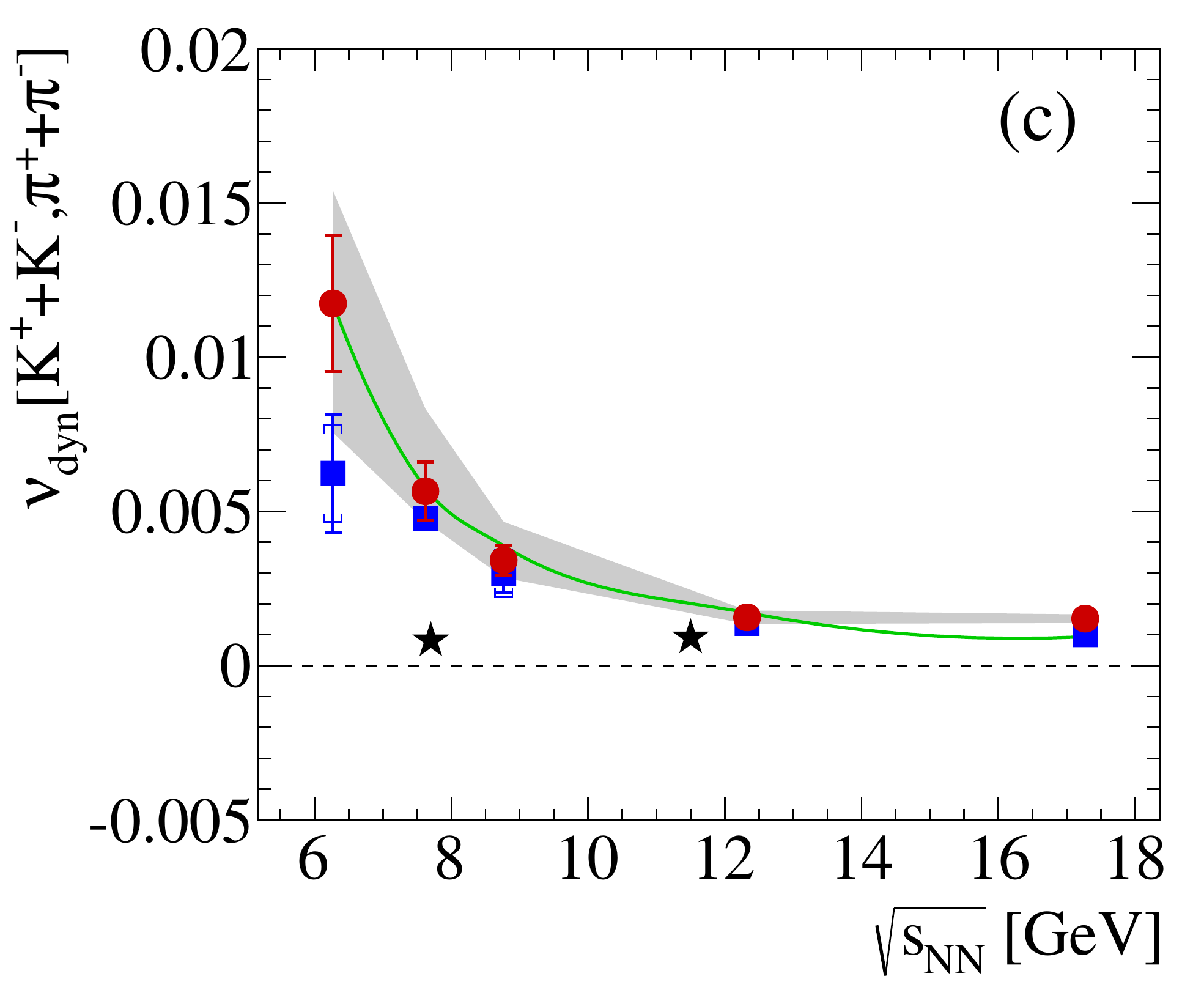}
\end{center}
\caption{(Color Online) Energy dependence of (a) $\nu_{\text{dyn}}[p+\bar{p},\pi^{+}+\pi^{-}]$, 
(b) $\nu_{\text{dyn}}[K^{+}+K^{-},p+\bar{p}]$ and (c) $\nu_{\text{dyn}}[K^{+}+K^{-},\pi^{+}+\pi^{-}]$. 
Results from the \textit{Identity Method} for central Pb+Pb data of NA49 are shown by red solid circles. 
Published NA49 results, converted from $\sigma_{\text{dyn}}$ to $\nu_{\text{dyn}}$ using Eq.~(\ref{relation}), 
are indicated by blue squares. Stars represent results 
of the STAR collaboration for central Au+Au collisions. In addition, for cases (a) and (c), 
the energy dependence predicted by Eq.(\ref{nuScaling}) is displayed by the green curves, 
which are consistent with the experimentally established trend. 
The systematic errors (see sections~\ref{sec-errors} and~\ref{sec-nu}) are presented as shaded bands.  
}
\label{nu_dyn}
\end{figure}

 \begin{center}
 \begin{table*}
\begin{tabular}{|c|c|c|c|}
\hline 
 & $\nu_{\text{dyn}}\times 1000$ & $\sigma_{\text{stat.}}\times 1000$ & $\sigma_{\text{sys.}}\times 1000$\\ [0.5ex] 
\hline  
20\textit{A} GeV & -6.139 & $\pm$ 0.243 & $^{+ 0.251}_{ - 0.190}$ \\  [0.5ex] 
\hline  
30\textit{A} GeV & -5.282 & $\pm$ 0.191 & $^{+ 0.206}_{ - 0.126}$ \\  [0.5ex] 
\hline  
40\textit{A} GeV & -5.058 & $\pm$ 0.125 & $^{+ 0.160}_{ - 0.068}$ \\  [0.5ex] 
\hline  
80\textit{A} GeV & -4.361 & $\pm$ 0.134 & $^{+ 0.346}_{ - 0.235}$ \\  [0.5ex]
\hline  
160\textit{A} GeV & -2.706 & $\pm$ 0.329 & $\pm$ 0.025 \\  [0.5ex] 
\hline  
 \end{tabular}
\caption{Numerical values of  $\nu_{\text{dyn}}[p+\bar{p},\pi^{+}+\pi^{-}]\times 1000$ with statistical and
systematic error estimates.}
\label{tableNuPPi}
\end{table*}
\end{center}

\begin{center}
 \begin{table*}
\begin{tabular}{|c|c|c|c|}
\hline 
 & $\nu_{\text{dyn}}\times 1000$ & $\sigma_{\text{stat.}}\times 1000$ & $\sigma_{\text{sys.}}\times 1000$\\ [0.5ex] 
\hline  
20\textit{A} GeV & 6.503 & $\pm$ 2.226 & $^{+ 3.808}_{ - 4.92}$ \\  [0.5ex] 
\hline  
30\textit{A} GeV & 2.210 & $\pm$ 1.122 & $^{+ 2.985}_{ - 1.099}$ \\  [0.5ex] 
\hline  
40\textit{A} GeV & -0.949 & $\pm$ 0.759 & $^{+ 1.422}_{ - 0.693}$ \\  [0.5ex] 
\hline  
80\textit{A} GeV & -2.498 & $\pm$ 0.587 & $^{+ 0.513}_{ - 0.099}$ \\  [0.5ex]
\hline  
160\textit{A} GeV & -2.135 & $\pm$ 0.460 & $\pm$ 0.001 \\  [0.5ex] 
\hline  
 \end{tabular}
\caption{Numerical values of  $\nu_{\text{dyn}}[K^{+}+K^{-},p+\bar{p}]\times 1000$ with statistical and
systematic error estimates.}
\label{tableNuPK}
\end{table*}
\end{center}

\begin{center}
 \begin{table*}
\begin{tabular}{|c|c|c|c|}
\hline 
 & $\nu_{\text{dyn}}\times 1000$ & $\sigma_{\text{stat.}}\times 1000$ & $\sigma_{\text{sys.}}\times 1000$\\ [0.5ex] 
\hline  
20\textit{A} GeV & 11.738 & $\pm$ 2.207 & $^{+ 3.647}_{ - 4.183}$ \\  [0.5ex] 
\hline  
30\textit{A} GeV & 5.651 & $\pm$ 0.943 & $^{+ 2.672}_{ - 0.972}$ \\  [0.5ex] 
\hline  
40\textit{A} GeV & 3.41816 & $\pm$ 0.485 &$^{ + 1.241}_{ - 0.569}$ \\  [0.5ex] 
\hline  
80\textit{A} GeV & 1.564 & $\pm$ 0.322 & $^{+ 0.225}_{ - 0.212}$ \\  [0.5ex]
\hline  
160\textit{A} GeV & 1.523 & $\pm$ 0.257 & $\pm$ 0.139 \\  [0.5ex] 
\hline  
 \end{tabular}
\caption{Numerical values of  $\nu_{\text{dyn}}[K^{+}+K^{-},\pi^{+}+\pi^{-}]\times 1000$ with statistical and
systematic error estimates.}
\label{tableNuKPi}
\end{table*}
\end{center}

The measure $\nu_{\text{dyn}}[A,B]$ of dynamical event-by-event fluctuations of the particle composition 
is defined as~\cite{voloshin_nu}:
\begin{equation}
\nu_{\text{dyn}}[A,B]=\frac{\left<A(A-1)\right>}{\left<A\right>^{2}} + \frac{\left<B(B-1)\right>}{\left<B\right>^{2}} - 2\frac{\left<AB\right>}{\left<A\right>\left<B\right>} , 
\label{eqNuDef1}
\end{equation}
where A and B stand for multiplicities of different particle species. As seen from the definition, Eq.(\ref{eqNuDef1}), the value of
$\nu_{\text{dyn}}$ vanishes when the multiplicity distributions of particles A and B follow the Poisson distribution and when there 
are no correlations between these particles ($\langle AB \rangle = \langle A \rangle \langle B \rangle$). On the other hand, a positive correlation term reduces the value 
of $\nu_{\text{dyn}}$, while an anticorrelation increases it. Inserting the values of 
the reconstructed moments (see Ref.~\cite{EDMS} for precise values)
into Eq.(\ref{eqNuDef1}) one obtains the values of $\nu_{\text{dyn}}[p+\bar{p}, \pi^{+}+\pi^{-}]$, $\nu_{\text{dyn}}[K^{+}+K^{-},p+\bar{p}]$ 
and  $\nu_{\text{dyn}}[K^{+}+K^{-},\pi^{+}+\pi^{-}]$. These results are represented by red solid circles in  Fig.~\ref{nu_dyn}.  
Statistical errors $\sigma_{\text{stat}}$ were estimated using the subsample method discussed in section~\ref{sec-errors}. 
Systematic uncertainties due to the applied track selection criteria were estimated by calculating $\nu_{\text{dyn}}$ 
separately for tracks selected by "loose" ($\nu_{\text{dyn}}^{\text{loose}}$) and "tight" ($\nu_{\text{dyn}}^{\text{tight}}$) 
cuts, while the systematic errors stemming from the uncertainty of the kaon fit were estimated using the K-S test (see section~\ref{sec-cuts}). 
The shift values of the fitted kaon positions, obtained from the K-S test for each phase-space bin, 
were used to obtain the values of $\nu_{\text{dyn}}^{\text{up}}$ and $\nu_{\text{dyn}}^{\text{down}}$. 
Final results (red solid circles in Fig.~\ref{nu_dyn}) are then presented as:

\begin{equation}
\nu_{\text{dyn}}[A,B]=\frac{\nu_{\text{dyn}}^{loose}+\nu_{\text{dyn}}^{tight}}{2},
\label{eqNuDef}
\end{equation}
the statistical errors are estimated using the Eq.~\ref{stat-1}, while the systematic errors, presented with shaded areas in Fig.~\ref{nu_dyn} are calculated as:

\begin{equation}
\sigma_{\text{sys}}^{\text{k}}=\text{sgn}\left(\nu_{\text{dyn}}^{k} - \nu_{\text{dyn}}\right)\sqrt{\left(\nu_{\text{dyn}}^{k} - \nu_{\text{dyn}}\right)^2 + 
\left(\frac{\nu_{\text{dyn}}^{loose}-\nu_{\text{dyn}}^{tight}}{2}\right)^2}.
\label{eqNuDef3}
\end{equation}
with k=(up, down).

These results (see Fig.~\ref{nu_dyn} and Tables~\ref{tableNuPPi},~\ref{tableNuPK} and~\ref{tableNuKPi}) are consistent 
with the values of $\nu_{\text{dyn}}$ obtained via Eq.~\ref{relation}  from the previously published NA49 measurements 
of the related measure 
$\sigma_{\text{dyn}}$~\cite{NA49_fluct1,NA49_fluct2} (blue squares in Fig.~\ref{nu_dyn}). Note that the source of systematic errors due to the uncertainties in kaon position were not considered in
previously published NA49 results, hence the presented systematic errors (blue horizontal bars) were underestimated.  We thus conclude that 
the increasing trend of the excitation functions of $\nu_{\text{dyn}}[K^{+}+K^{-},p+\bar{p}]$ and $\nu_{\text{dyn}}[K^{+}+K^{-},\pi^{+}+\pi^{-}]$ 
towards low energies is confirmed by two independent analyses of the NA49 data on 
central Pb+Pb collisions. Also presented in Fig.~\ref{nu_dyn} 
are the STAR results (black stars) from the RHIC Beam Energy Scan (BES) program~\cite{STAR_fluct} 
for central Au+Au collisions, which clearly differ at low energies. However, as mentioned above, 
the phase space coverage of NA49 and STAR are not the same. The consequences will be discussed below. 

\begin{figure}[htbp]
\begin{center}
\includegraphics[width=0.32\textwidth]{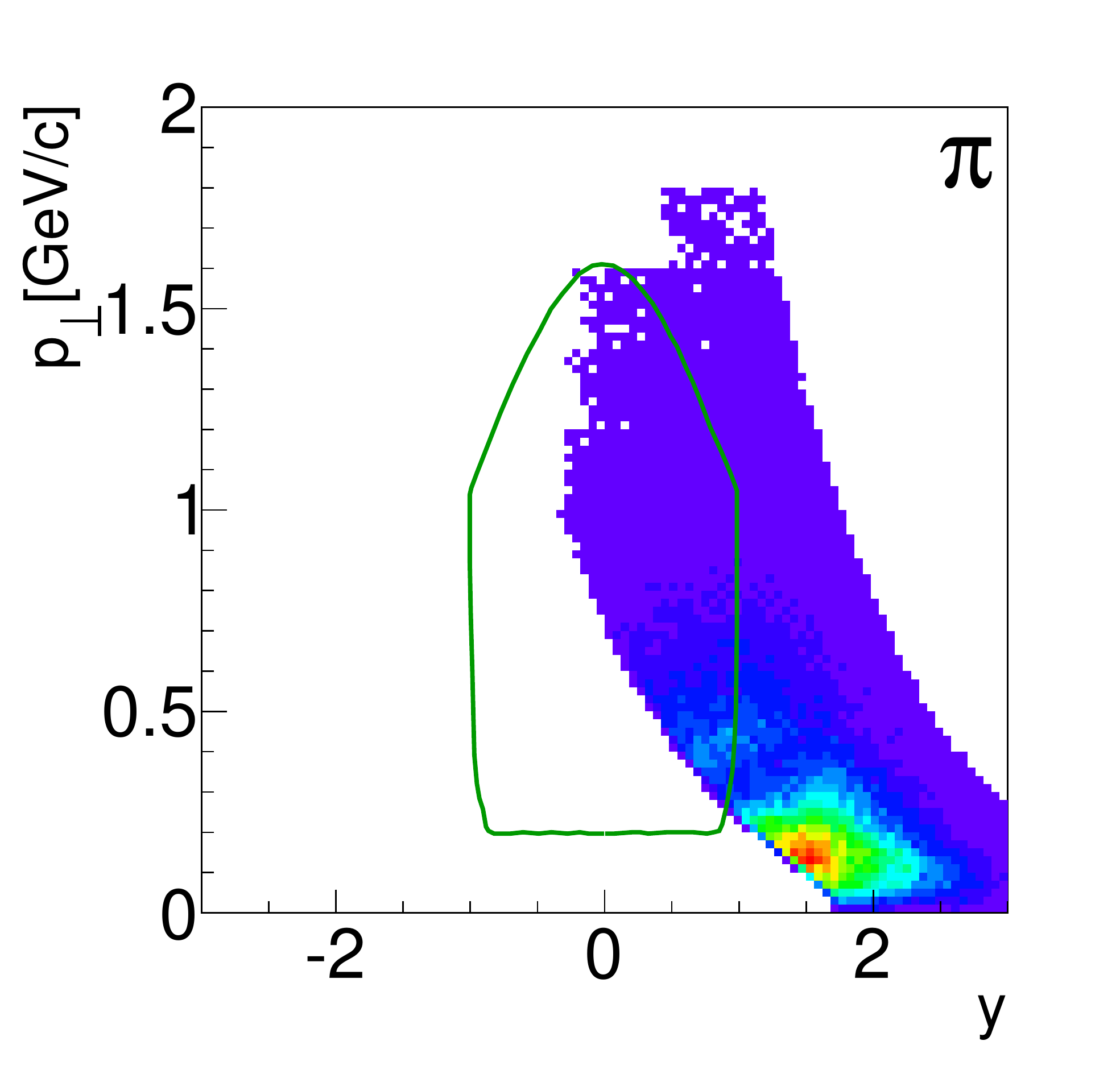}
\includegraphics[width=0.32\textwidth]{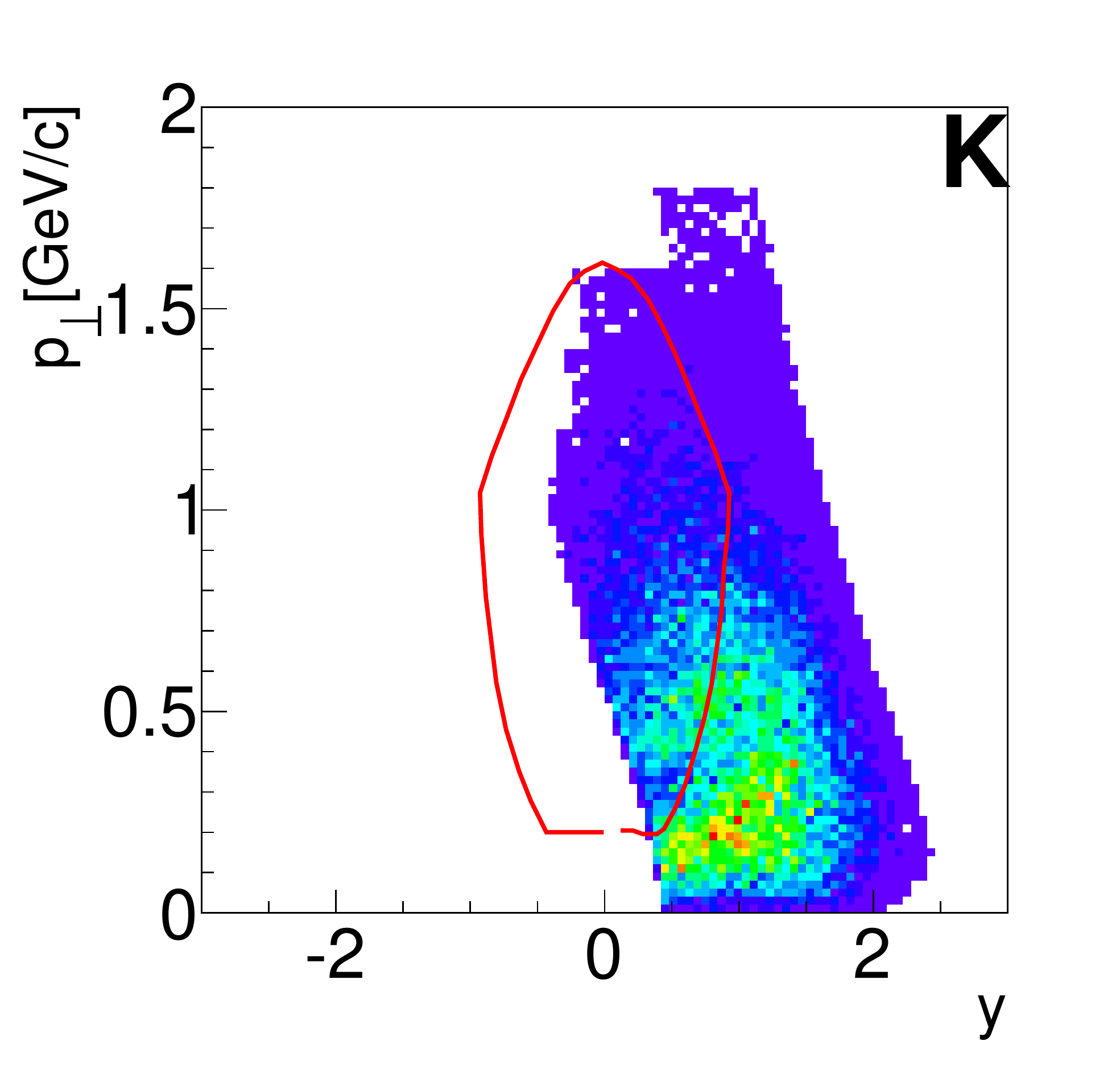}
\includegraphics[width=0.32\textwidth]{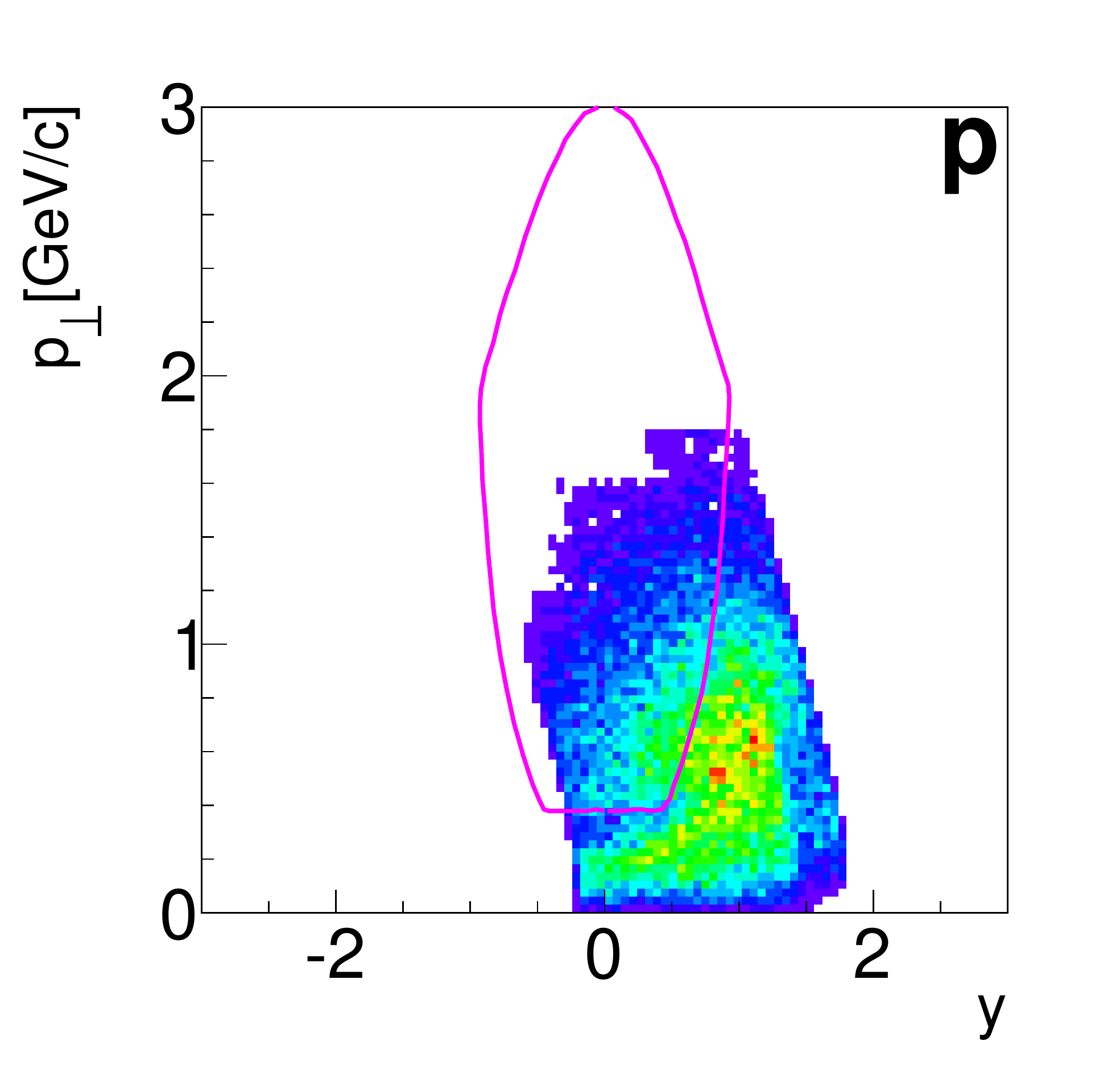}
\includegraphics[width=0.32\textwidth]{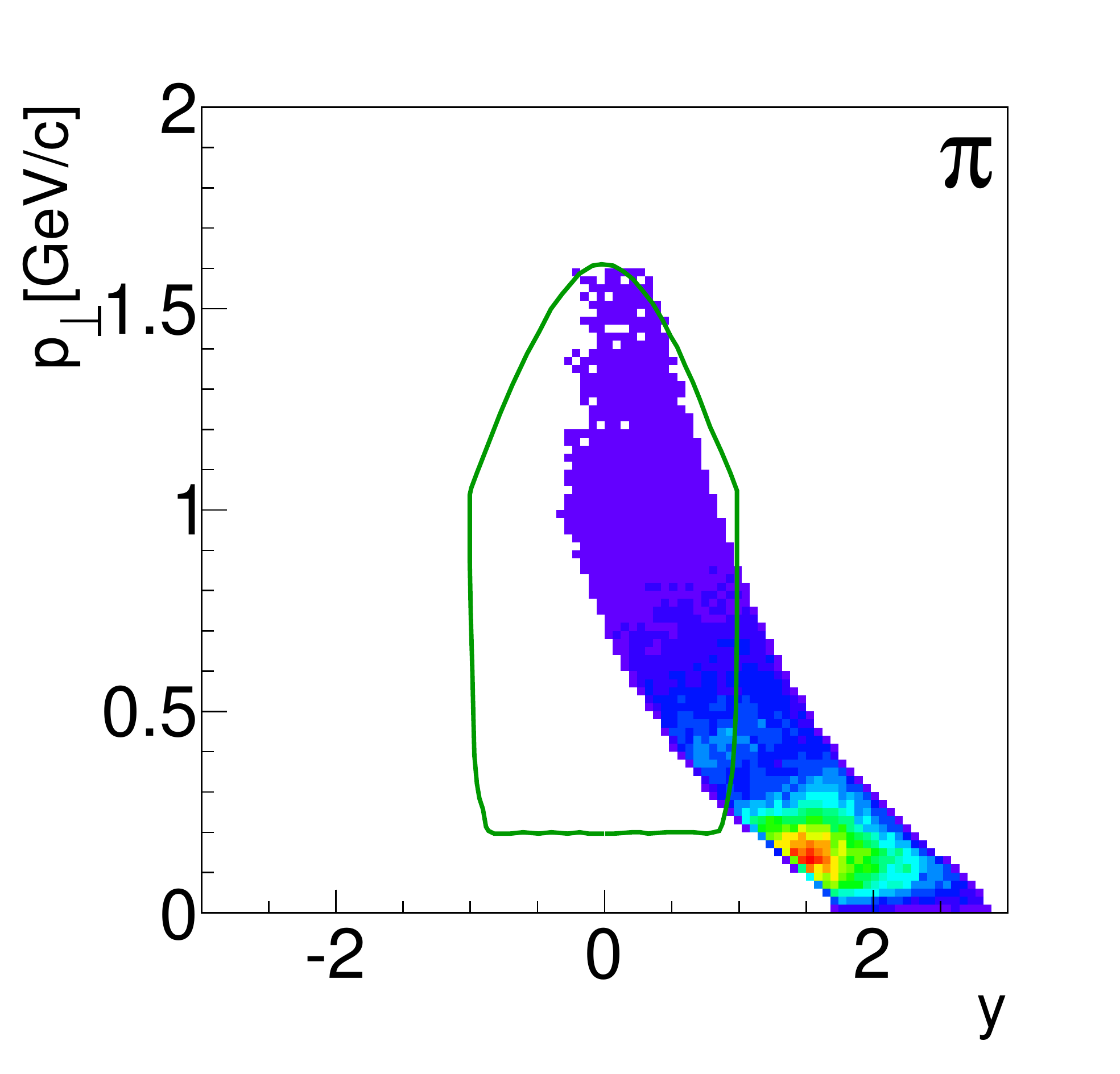}
\includegraphics[width=0.32\textwidth]{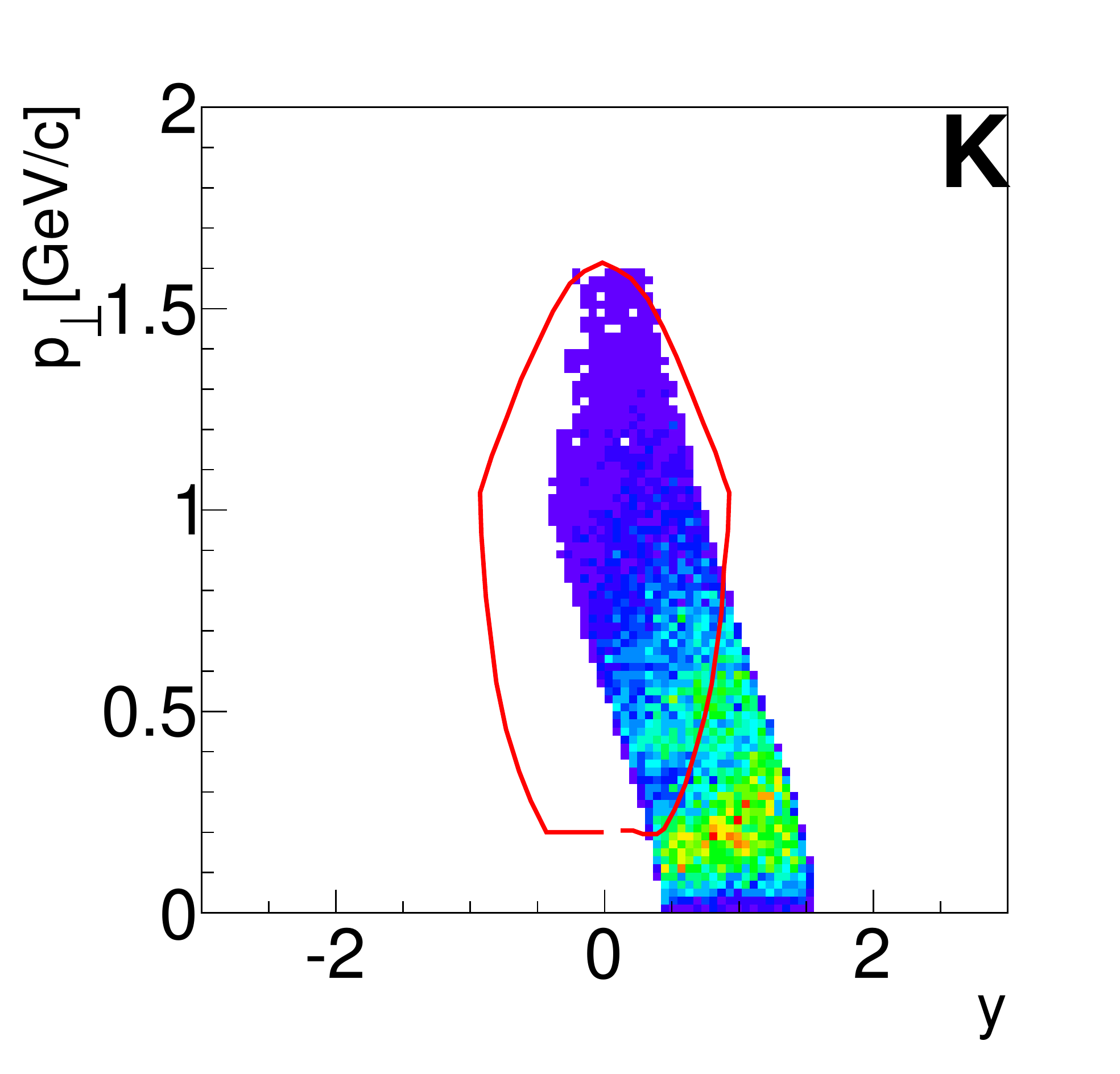}
\includegraphics[width=0.32\textwidth]{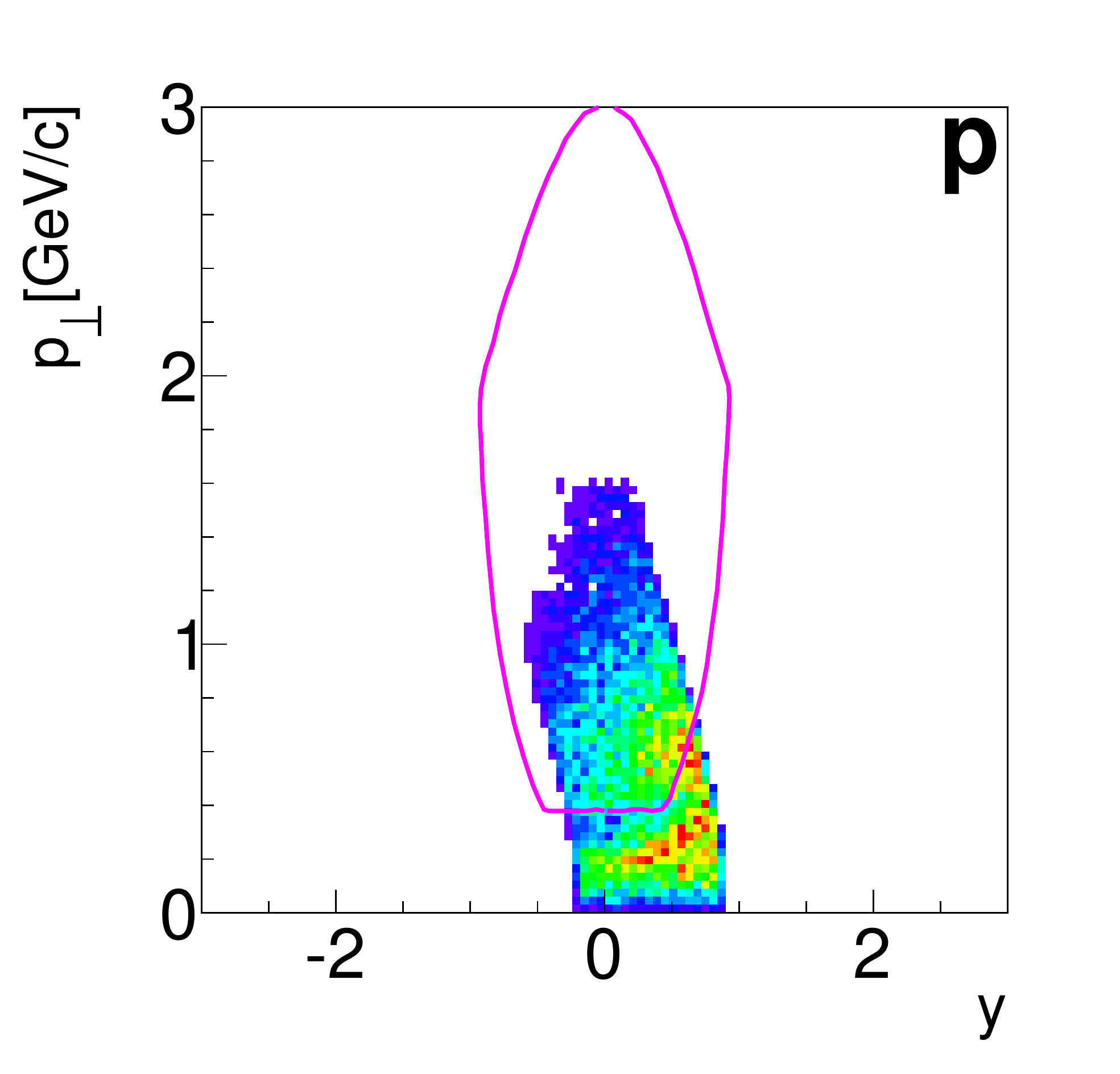}
\end{center}
\caption{(Color Online) Phase-space coverage for identified pions, kaons and protons 
in the acceptance of the NA49 experiment for Pb+Pb collisions at 30\textit{A}~GeV/c 
(upper panels). Lower panels illustrate an example of a restriction of the
phase-space coverage to better match the region covered by STAR (indicated by solid lines) 
at the corresponding beam energy.
}
\label{phase_space}
\end{figure}


\section{Phase space dependence of $\nu_{\text{dyn}}$ measurements}
\label{sec-accepatance}

\begin{figure}[htbp]
\begin{center}
\includegraphics[width=0.5\textwidth]{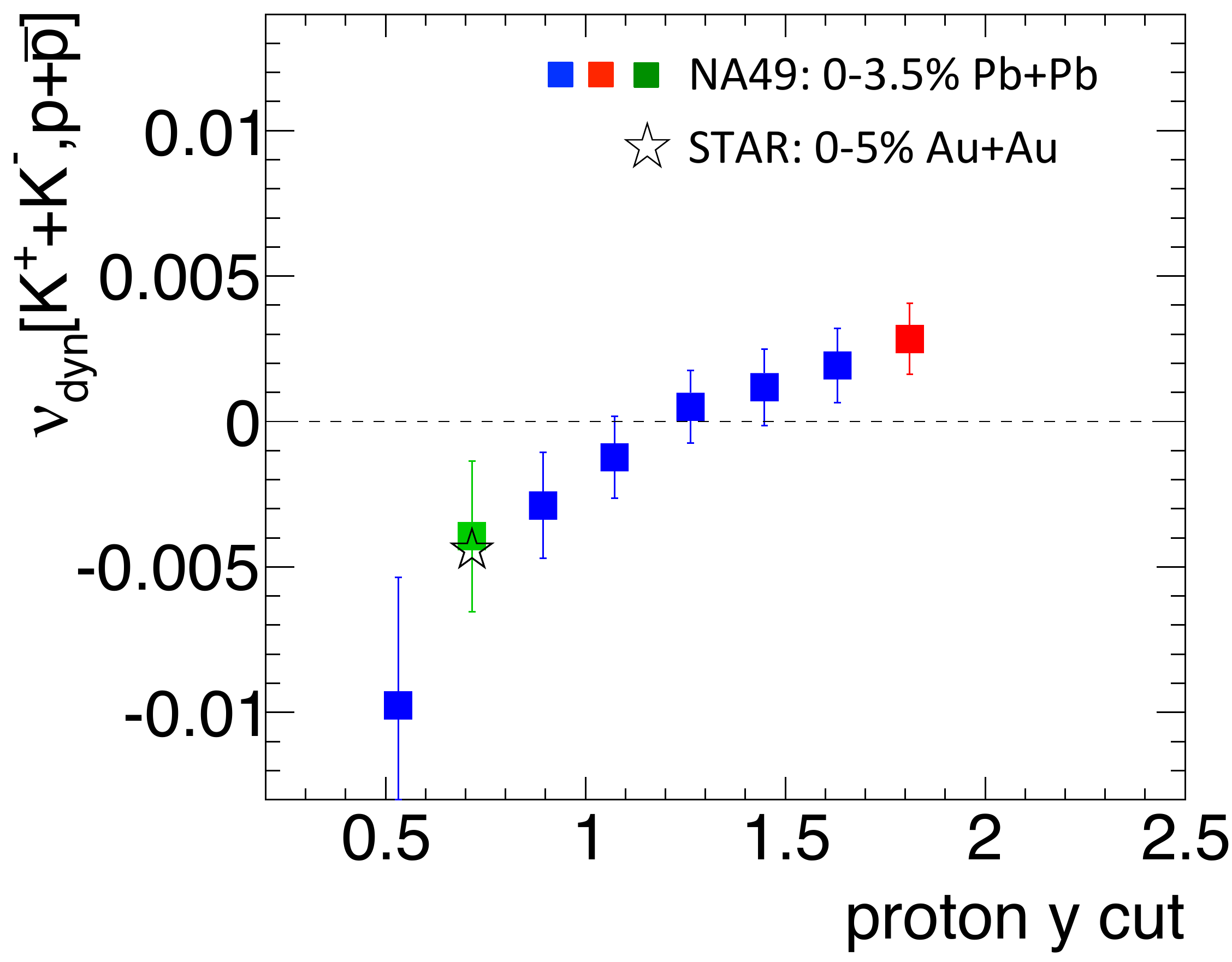}
\end{center}
\caption{(Color Online) Phase space dependence of $\nu_{\text{dyn}}[K^{+}+K^{-},p+\bar{p}]$ for 30\textit{A} GeV 
Pb+Pb collisions of NA49. Red and green squares correspond to the phase space bins illustrated in the 
upper and lower panels of Fig.~\ref{phase_space} respectively. Blue squares are the NA49 results 
for other phase space bins. The result of the STAR experiment is plotted as the purple star
at the corresponding NA49 phase space bin. The phase space region of the analysis is varied by
an upper cut on the momentum (see text).
}
\label{explain}
\end{figure}

\begin{figure}[htbp]
\begin{center}
\includegraphics[width=0.45\textwidth]{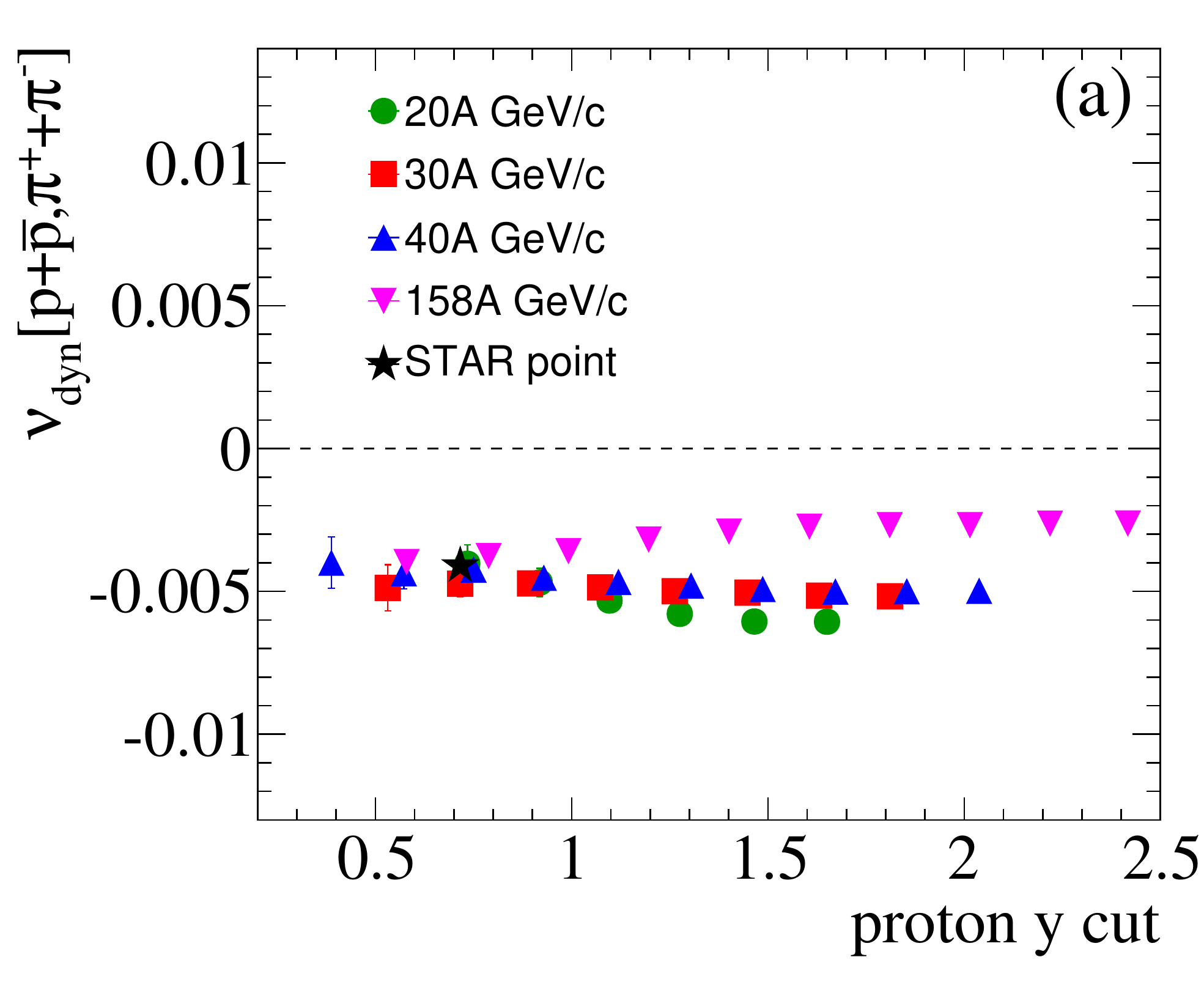}
\includegraphics[width=0.45\textwidth]{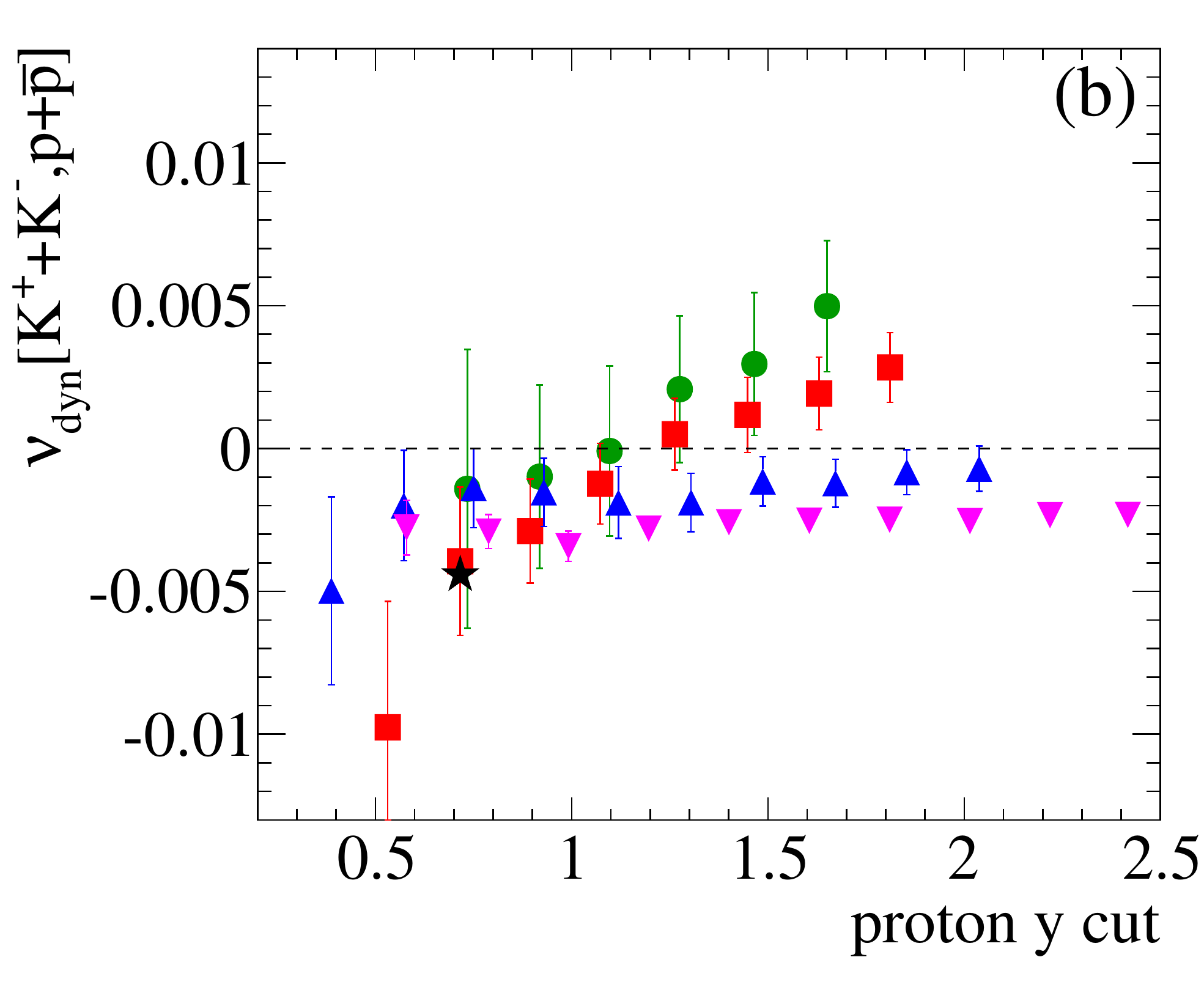}
\includegraphics[width=0.45\textwidth]{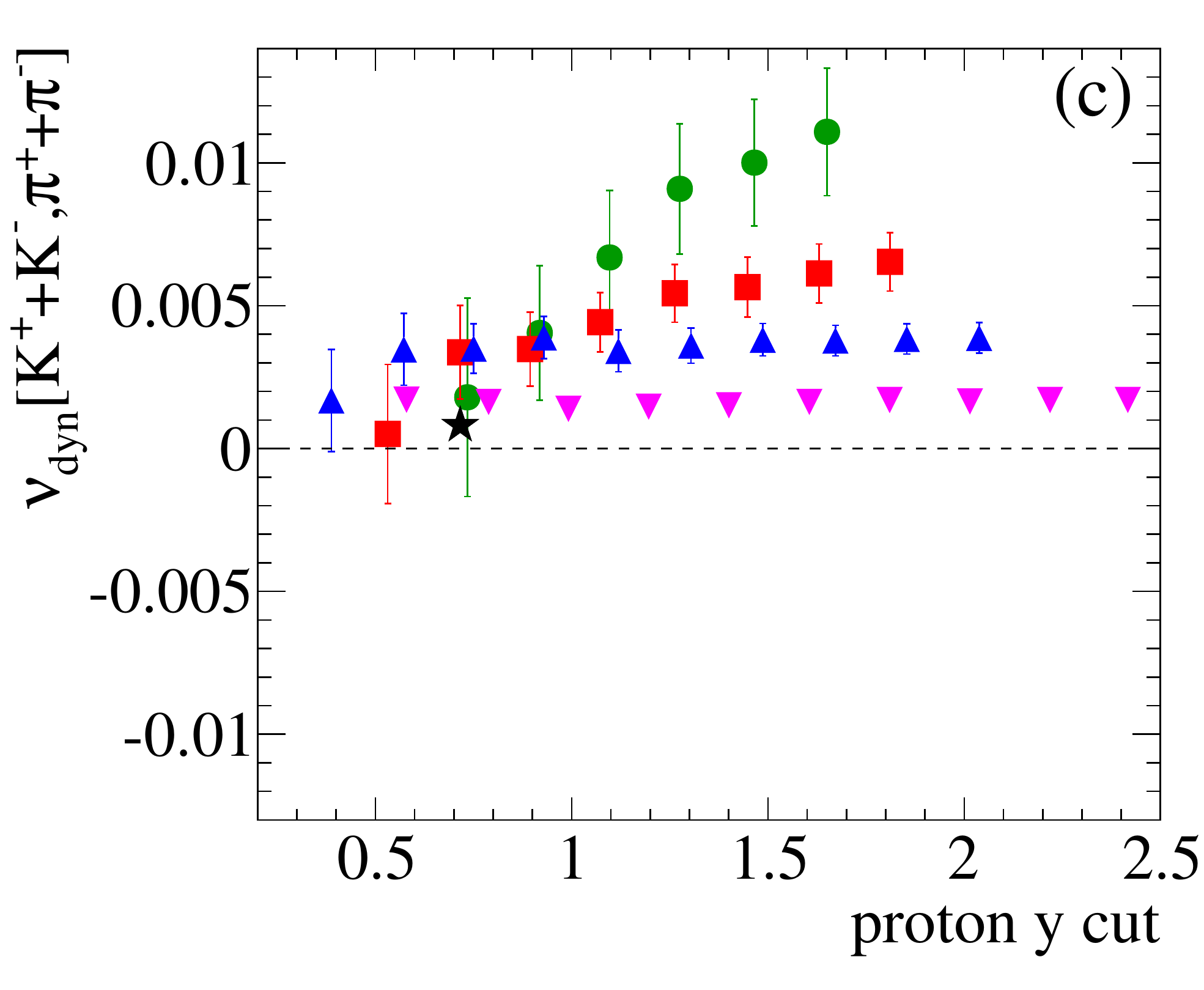}
\end{center}
\caption{(Color Online) Phase-space region dependence of (a) $\nu_{\text{dyn}}[p+\bar{p},\pi^{+}+\pi^{-}]$, 
(b) $\nu_{\text{dyn}}[K^{+}+K^{-},p+\bar{p}]$ and (c) $\nu_{\text{dyn}}[K^{+}+K^{-},\pi^{+}+\pi^{-}]$ 
in central Pb+Pb collisions of NA49 (triangles, squares, dots).
Stars show measurements of the STAR collaboration.
Results are plotted versus the maximum proton rapidity (see text). 
}
\label{acc_dep}
\end{figure}

The investigation presented in this section attempts to shed light on the cause of the differences between 
the results from STAR and NA49 on fluctuations of identified hadrons. Two sources were studied: the dependence
of $\nu_{\text{dyn}}$ on the multiplicity of the particles entering the analysis and a possible sensitivity
of $\nu_{\text{dyn}}$ to the covered phase space region.

Indeed, it was found in Ref.~\cite{Tim_Volker} that $\nu_{\text{dyn}}$  exhibits an intrinsic dependence on the 
multiplicities of accepted particles. Since multiplicities increase with increasing collision energy, this leads to 
a trivial energy dependence of $\nu_{\text{dyn}}$:

\begin{equation}
\nu_{\text{dyn}}[A,B](E)=  \nu_{\text{dyn}}[A,B](E_{ref})\frac{\left[\frac{1}{\langle A \rangle}+\frac{1}{ \langle B \rangle }\right]_{E}}{\left[\frac{1}{\langle A\rangle }+\frac{1}{\langle B \rangle}\right]_{E_{ref}}},
\label{nuScaling}
\end{equation}
where $E_{ref}$ is the energy at which the reference value of $\nu_{\text{dyn}}$ was chosen and the $E$ denotes the energy at which the value of $\nu_{\text{dyn}}$ is estimated. The 
energy dependence predicted by Eq.(\ref{nuScaling}), with a reference energy of $E_{ref} = \sqrt{s_{NN}}\approx$~6.3 GeV 
(corresponding to 20$\mathit{A}$ GeV laboratory energy), is illustrated for $\nu_{\text{dyn}}[p+\bar{p}, \pi^{+}+\pi^{-}]$ 
and $\nu_{\text{dyn}}[K^{+}+K^{-},\pi^{+}+\pi^{-}]$ in Fig.~\ref{nu_dyn}(a and c) by the green curves. However, this scaling
prescription cannot reproduce the sign change observed for the energy dependence of $\nu_{\text{dyn}}[K^{+}+K^{-},p+\bar{p}]$
as shown in Fig.~\ref{nu_dyn}(b). Moreover, using the multiplicities of Table~\ref{tableMoments} and the corresponding
numbers for the STAR experiment \cite{Tarnowsky_private} one would expect only about a factor 2 decrease of the value of
$\nu_{\text{dyn}}[K^{+}+K^{-},\pi^{+}+\pi^{-}]$ at $\sqrt{s_{NN}}=7.6$ GeV which does not lead to agreement with the STAR result.

Next, the sensitivity of $\nu_{\text{dyn}}$ to the covered regions of phase space will be studied since these differ for
the NA49 and STAR measurements.
As an example Fig.~\ref{phase_space} illustrates the phase space coverage for pions, kaons and protons 
at 30$A$~GeV projectile energy in the acceptance of the NA49 detector. In the same figure the acceptance 
of the STAR apparatus at corresponding center-of-mass energy is presented by colored lines. 
The dependence of $\nu_{\text{dyn}}$ on the selected phase space region was studied by performing the analysis in different 
phase space bins stretching from a forward rapidity cut to mid-rapidity. Technically different 
phase space bins were selected by applying upper momentum cuts to the reconstructed tracks 
where the cut value corresponded to the momentum of a proton at $p_{\bot}$=0 with 
a chosen maximum rapidity. Thereafter this quantity will be called a proton rapidity cut. 
The upper panels of Fig.~\ref{phase_space} illustrate one such phase space bin for 30\textit{A}~GeV Pb+Pb data. 
The reconstructed value of $\nu_{\text{dyn}}[K^{+}+K^{-},p+\bar{p}]$ in this bin is plotted as a red square in 
Fig.~\ref{explain}. Similarly the green square in Fig.~\ref{explain} represents the reconstructed value 
of $\nu_{\text{dyn}}[K^{+}+K^{-},p+\bar{p}]$ corresponding to the phase space bin plotted in the lower panel 
of Fig.~\ref{phase_space}. Note that in this particular bin the NA49 point is consistent with the STAR 
result, which is shown by the purple star. This study demonstrates a strong dependence of the 
resulting value of $\nu_{\text{dyn}}$ on the phase space covered by the measurement. 
Fig.~\ref{acc_dep} shows the dependence of $\nu_{\text{dyn}}$ for different combinations of particles at 
different energies. At 20\textit{A} and 30\textit{A}~GeV $\nu_{\text{dyn}}[K^{+}+K^{-},p+\bar{p}]$ and 
$\nu_{\text{dyn}}[K^{+}+K^{-},\pi^{+}+\pi^{-}]$ show a strong dependence on the extent of the phase space region
and eventually hit the 
STAR point in a particular bin. Interestingly the acceptance dependence  weakens above 30\textit{A}~GeV 
where no difference was observed with STAR. It is also remarkable that $\nu_{\text{dyn}}[p+\bar{p},\pi^{+}+\pi^{-}]$ 
shows little dependence on the covered phase space region. 
This detailed study of $\nu_{\text{dyn}}$ in different phase space regions appears to explain 
to a large extent the difference between the STAR BES and NA49 measurements.

Some final remarks are in order concerning the properties and the significance of the fluctuation
measure $\nu_{\text{dyn}}$.
To reveal the physics underlying the studied event-by-event fluctuations, the fluctuation 
signals measured in heavy-ion (A+A) collisions should be compared systematically to a reference from 
nucleon-nucleon (N+N) collisions at corresponding energies per nucleon. It is however important 
to properly take into account trivial differences between A+A and N+N collisions e.g. 
in the size of the colliding systems.
An additional complication in the experimental study of fluctuations in A+A collisions are 
unavoidable volume fluctuations from event to event. To take account of these considerations a
set of "strongly intensive" fluctuation measures has been proposed in Ref.~\cite{S_I}. In fact, the scaled $\nu_{\text{dyn}}$ (see Eq.(\ref{nuScaling})) is related to the
strongly intensive measure $\Sigma^{\text{AB}}$ (cf. Eq.(13) in Ref.~\cite{S_I}):

\begin{equation}
\nu_{\text{dyn}}[A,B]^{Scaled} \equiv \frac{\nu_{\text{dyn}}[A,B]}{\frac{1}{\langle A \rangle}+\frac{1}{\langle B \rangle}} = \Sigma^{AB} -1.
\label{SigmaDef}
\end{equation}

Future studies of strongly intensive measures may lead to a better understanding of the underlying source of correlations.
 
%
%
%
%
%

\section{Summary}
\label{sec-summary}
In summary several scenarios were investigated to understand the differences between the NA49 and STAR 
measurements of the excitation functions of $\nu_{\text{dyn}}[K^{+}+K^{-},p+\bar{p}]$ and $\nu_{\text{dyn}}[K^{+}+K^{-},\pi^{+}+\pi^{-}]$.  
For this purpose the particle identification procedure formerly employed by NA49 was replaced by
a different approach, the \textit{Identity Method}, to reconstruct the fluctuation measure $\nu_{\text{dyn}}$. 
The increasing trend of  $\nu_{\text{dyn}}[K^{+}+K^{-},p+\bar{p}]$ and $\nu_{\text{dyn}}[K^{+}+K^{-},\pi^{+}+\pi^{-}]$ towards lower energies reported 
in previous publications of NA49 in terms of the quantity $\sigma_{\text{dyn}}$ was confirmed by this analysis.
A detailed study of $\nu_{\text{dyn}}$ reveals a strong dependence on the phase space coverage at low energies 
for $\nu_{\text{dyn}}[K^{+}+K^{-},p+\bar{p}]$ and $\nu_{\text{dyn}}[K^{+}+K^{-},\pi^{+}+\pi^{-}]$ which might explain the different energy dependences
measured by NA49 (central Pb+Pb collisions) and STAR (BES program for central Au+Au collisions).
As an outlook it is worth mentioning that since the \textit{Identity Method} reconstructs
first and second moments of the multiplicity distributions of identified particles
one will be able to investigate the energy dependence of
all the fluctuation measures proposed in Ref.~\cite{S_I}.
These quantities are better suited for phase transition studies because (within the grand canonical ensemble)
they depend neither on the volume nor on its fluctuations which cannot be tightly controlled in
experiments.

\vspace{1.0cm}
{\bf Acknowledgments:}~This work was supported by
the US Department of Energy Grant DE-FG03-97ER41020/A000,
the Bundesministerium fur Bildung und Forschung (06F 137), Germany,
the German Research Foundation (grant GA 1480/2.1),
the National Science Centre, Poland (grants DEC-2011/03/B/ST2/02617 and DEC-2011/03/B/ST2/02634),
the Hungarian Scientific Research Foundation (Grants OTKA 68506, 71989, A08-77719 and A08-77815),
the Bolyai Research Grant,
the Bulgarian National Science Fund (Ph-09/05),
the Croatian Ministry of Science, Education and Sport (Project 098-0982887-2878)
and
Stichting FOM, the Netherlands.


\begin{thebibliography}{00} 
\bibitem{QGP1} J. C. Collins and M. J. Perry, Phys. Rev. Lett. 34, 1353 (1975).
\bibitem{QGP2} E. V. Shuryak, Phys. Rept. 61, 71 (1980), Phys. Rept. 115, 151 (1984).
\bibitem{LatticeQCD} for recent results see: C.~Schmidt (RBC-Bielefeld and HotQCD Collaborations),
   Nucl.~Phys.~A820, 41c (2009); Z.~Fodor and S.~Katz (Wuppertal Collaboration), Acta~Phys.~Pol.~B42, 2791 (2011).
\bibitem{Stephanov} M.~Stephanov, Int.~J.~Mod.~Phys.~A20, 4387 (2005).
\bibitem{StepRajShur} M.~Stephanov, K.~Rajagopal and E.~Shuryak, Phys. Rev. D60, 114028 (1999).
\bibitem{Koch} V.~Koch, arXiv:0810.2520 (2008).
\bibitem{HeinzJacob} U.~Heinz and M.~Jacob, arXiv:nucl-th/0002042 (2000).
\bibitem{SMES} M.~Gazdzicki and M.~Gorenstein, Acta~Phys.~Pol.~B30, 2705 (1999).
\bibitem{NA49_horn} C. Alt et al. (NA49 Collab.), Phys. Rev. C77, 024903 (2008).
\bibitem{NA49_Rustamov} A. Rustamov, Central Eur. J. Phys. 10, 1267-1270 (2012), arXiv:1201.4520v1 [nucl-ex] (2012).
\bibitem{NA49_fluct1} C. Alt et al. (NA49 Collab.), Phys. Rev. C79, 044910 (2009).
\bibitem{NA49_fluct2} T. Anticic et al. (NA49 Collab.) Phys. Rev. C83, 061902(R) (2011).
\bibitem{NA49_fluct3} T. Anticic et al. (NA49 Collab.), Phys. Rev. C87, 024902 (2013).
\bibitem{voloshin_nu} C. Pruneau, S. Gavin, S. Voloshin, Phys. Rev. C66, 044904 (2002).
\bibitem{STAR_fluct} T.~Tarnowsky (STAR Collab.) J. Phys. G: Nucl. Part. Phys. 38 124054 (2011).
\bibitem{identity1} M. Gazdzicki et al., Phys. Rev. C83, 054907 (2011).
\bibitem{identity2} M. I. Gorenstein, Phys. Rev. C84, 024902 (2011).
\bibitem{NA49_NIM} S. Afanasiev et al. (NA49 Collab.), Nucl. Instrum. Meth. A430, 210 (1999).
\bibitem{glauber}  K. J. Eskola, K. Kajantie and J. Lindfors, Nucl. Phys. B323, 37 (1989)
\bibitem{dedx_Landau} L. Landau, Journal of Physics (USSR), vol. 8, p. 201 (1944).
\bibitem{dedx_Marco} M. van Leeuwen, PhD thesis, NIKHEFF, Amsterdam (2003), CERN EDMS Id 816033;
NA49 technical notes G. Veres, CERN EDMS Id 815871 (2000), M. van Leeuwen, CERN EDMS Id 983015.
\bibitem{EDMS} \url{https://edms.cern.ch/file/1311963/1/NA49_paper_moments.pdf}.
\bibitem{identity3} A. Rustamov, M. I. Gorenstein, Phys. Rev. C86, 044906 (2012).
\bibitem{Tim_Volker} V. Koch and T. Schuster, Phys. Rev. C81, 034910 (2010).
\bibitem{Tarnowsky_private} T.~Tarnowsky (STAR Collab.), private communication.
\bibitem{S_I} M. I. Gorenstein and M. Gazdzicki, Phys. Rev. C84,  014904 (2011).  
\end{thebibliography}
\end{document}